\documentclass[12pt]{article}

\usepackage{lineno,hyperref,amsmath,amsfonts,amssymb,amsthm,diagbox,graphicx}
\usepackage{caption, subcaption, color, natbib}
\usepackage{setspace}
\usepackage[left=1in,top=1in,right=1in,bottom=1in,nohead,paperwidth=8.5in, paperheight=11in]{geometry} 
\usepackage{authblk}

\newcommand{\bx}{\boldsymbol x}
\newcommand{\bz}{\boldsymbol z}
\newcommand{\bX}{\boldsymbol X}
\newcommand{\bD}{\boldsymbol D}
\newcommand{\bT}{\boldsymbol T}
\newcommand{\bU}{\boldsymbol U}
\newcommand{\bOmega}{\boldsymbol \Omega}
\newcommand{\bSigma}{\boldsymbol \Sigma}

\DeclareMathOperator*{\IF}{IF}
\DeclareMathOperator*{\SC}{SC}
\DeclareMathOperator{\E}{\mathbb{E}}
\DeclareMathOperator{\sign}{sign}
\DeclareMathOperator{\median}{median}
\DeclareMathOperator*{\vecmat}{vec}
\DeclareMathOperator*{\matvec}{mat}
\DeclareMathOperator*{\trace}{tr}
\DeclareMathOperator*{\argmin}{arg\,min}
\DeclareMathOperator*{\argmax}{arg\,max}

\newtheorem{remark}{Remark}
\newtheorem{theorem}{Theorem}
\newtheorem{lemma}{Lemma}

\title{\bf The Influence Function of Graphical Lasso Estimators}
\author[a]{Ga\"{e}tan Louvet\thanks{Corresponding author. Email: gaetan.louvet@unamur.be, postal address: Rue de Bruxelles, 61; 5000 Namur, Belgium}}
\author[b]{Jakob Raymaekers}
\author[a]{Germain Van Bever}
\author[b]{Ines Wilms}
\affil[a]{Department of Mathematics and Namur Institute for Complex Systems (Naxys), Universit\'{e} de Namur, Belgium}
\affil[b]{Department of Quantitative Economics,
Maastricht University, The Netherlands}
\date{\today}

\begin{document}
\onehalfspacing
\maketitle

\begin{abstract}
The precision matrix that encodes conditional linear dependency relations among a set of  variables 
forms an important object of interest in multivariate analysis. 
Sparse estimation procedures for precision matrices such as the graphical lasso (Glasso)  gained popularity as they facilitate interpretability, thereby sepa\-rating pairs of variables that are conditionally dependent from those that are independent (given all other variables).
Glasso lacks, however, robustness to outliers. To overcome this problem, one typically applies a robust plug-in procedure where the Glasso is computed from a robust covariance estimate instead of the sample covariance, thereby providing protection against outliers. 
In this paper, we study such estimators theoretically, by deriving and comparing their influence function, sensitivity curves and asymptotic variances. 
\bigskip


\end{abstract}
\doublespacing

\newpage
\section{Introduction} \label{sec:intro}
The estimation of precision (inverse covariance) matrices is  indispensable in multivariate analysis with applications ranging from  economics and finance to genomics, neuroscience and social networks  \citep{fan2014challenges}.
Let ${\bf X}=(X_1, \ldots, X_p)^\top$ be a $p$-variate random vector with finite second order moments, with mean $\boldsymbol\mu$ and covariance matrix $\bSigma$.
We focus on estimating the precision matrix $\bOmega:=\bSigma^{-1}$ whose entries capture the conditional linear dependencies between the components of ${\bf X}$.
Importantly, when ${\bf X}\sim N(\boldsymbol\mu, \bSigma)$, the variables $X_i$ and $X_j$ are conditionally independent given all other variables if and only if $(\bOmega)_{i,j} = 0$.
In that case, recovering the sparsity pattern of the precision matrix is equivalent to recovering the graph structure of the Gaussian graphical model $G=(V, E)$
where the vertex set $V=\{1, \ldots, p\}$ and the edge set $E$ consists of the pairs $(i,j)$ that are connected by an edge (thus having $(\bOmega)_{i,j} \neq 0$).

A common method for estimating sparse precision matrices is the graphical lasso  (Glasso) \citep{yuan2007model, banerjee2008model, rothman2008sparse, friedman2008sparse}, defined as 
\begin{equation}
\widehat{\bOmega} = \argmin_{\bOmega}
\{- \text{logdet}(\bOmega) + \text{tr}({\bf S}\bOmega) + \lambda \|\bOmega^{-\text{diag}}\|_1 \ \ \text{s.t.} \  \bOmega=\bOmega^\top, \bOmega \succ 0\}, \label{Glasso}
\end{equation}
where $\text{logdet}(\bOmega)=\log(|\bOmega|)$, $\text{tr}(\cdot)$ denotes the trace, $\bf S$ is the sample covariance matrix, $\bOmega^{-\text{diag}}$ is $\bOmega$ with each diagonal element set to $0$,  $\cdot \succ 0$ denotes a positive definite matrix, and $\lambda>0$ is a penalty parameter controlling the degree of sparsity.
Throughout, we refer to the estimator in equation \eqref{Glasso}, with the sample covariance matrix as input, as the ``standard Glasso". 
By adding an $\ell_1$-penalty (in this case, the $\ell_1$-norm of the off-diagonal elements of ${\bf \Omega}$)
to the negative log-likelihood of a sample of multivariate normal random variables, a sparse precision matrix estimate is obtained thereby also permitting estimation in high-dimensional settings with more parameters than observations. 

While sparse precision matrix estimation forms an active area of research, far fewer research has looked into the performance of such methods in the presence of outliers. 
Nonetheless, in many multivariate settings, the occurrence of 
outliers is to be expected. 
The standard Glasso, however, performs poorly in presence of outliers as it relies on the  sample covariance matrix ${\bf S}$ in equation \eqref{Glasso}, which is sensitive to even a single outlying data point.
Various robustification approaches of the Glasso have been proposed. 
\cite{finegold2011robust} build upon the standard Glasso but use the likelihood of the multivariate $t$-distribution instead of a normal one thereby providing better protection against data contamination. 
\cite{ollerer2015robust, croux2016robust, tarr2016robust, lafit2022robust} consider high-dimensional precision matrix estimation under cellwise contamination and propose to plug-in a pairwise robust covariance or correlation estimator  instead of the (non-robust) sample covariance into equation \eqref{Glasso}.
Applications of such robust sparse precision matrices to, for instance, discriminant analysis are studied in \cite{aerts2017cellwise}.
However, most studies above do not provide a theoretical analysis on robust Glasso estimators. 
Exceptions are \cite{ollerer2015robust,croux2016robust} who derive a finite sample breakdown point of their robust Glasso  under cellwise contamination, and \cite{loh2018high} who analyze the statistical consistency of robust high-dimensional precision matrix estimators under cellwise contamination.

In this paper, we  study the robustness of sparse Glasso-based precision matrix estimators theoretically by deriving their influence function.
The influence function is an important tool to measure the robustness of a statistical functional \citep{hampel86}; it quantifies the influence of a small amount of contamination placed at a given value $\bz$ on a statistical functional $T$, for data coming from a distribution $F$.
More precisely, the influence function of the statistical functional $T$ at the  distribution $F$ is, when it exists,
given by
\begin{equation}
\IF(\bz; T, F) = \underset{\varepsilon \rightarrow 0}{\text{lim}} \frac{T(F_{\varepsilon, \bz}) - T(F)}{\varepsilon} = \frac{\partial}{\partial\varepsilon}T(F_{\varepsilon, \bz})\Bigr|_{\varepsilon=0} ,
\end{equation}
where 
$F_{\varepsilon, z} = (1 - \varepsilon) F + \varepsilon \Delta(\bz)$ is the contaminated distribution and $\Delta(\bz)$ is the probability distribution that puts all its mass at $\bz$.
As commonly done in the study of influence functions, we thus limit ourselves to point mass contamination; different forms of outliers can be studied through other robustness concepts such as the breakdown point (see e.g., \cite{ollerer2015robust,croux2016robust}).

The remainder of the article is structured as follows. 
In Section \ref{sec:IF-Glasso}, we present the influence function of the Glasso for any  plug-in scatter functional, then discuss the special case of the standard Glasso and  show its influence function is unbounded.
Section \ref{sec:IF-robust} discusses how the Glasso influence function can be bounded and expressions of the gross-error sensitivity are derived.
We consider different robust Glasso estimators and compare them in terms of their influence functions and sensitivity curves.
In Section \ref{sec:ASV}, we compute the asymptotic variances and compare the robust Glasso estimators in terms of their statistical efficiencies.
Section \ref{sec:conclusion} concludes.
All proofs, together with precise descriptions of the robust correlation measures used in Section~\ref{sec:IF-robust}, are collected in the Appendix.

\section{Influence Function of the Glasso} \label{sec:IF-Glasso}
We first derive the general expression of the influence function of the Glasso for any plug-in scatter functional (Section \ref{subsec:IFGlasso}).
Next, we discuss the case of the standard Glasso and establish its lack of robustness (Section \ref{subsec:Glasso-nonrob}).

\subsection{The Influence Function} \label{subsec:IFGlasso}
We start with deriving the influence function of the associated functional representation of the Glasso.
The following notation will be used throughout. Let $p$ be fixed.
For any $p\times p$ matrix $\boldsymbol{\Omega}$, let $\boldsymbol{\omega}:=\vecmat(\boldsymbol{\Omega})$ denote the $p^2$-dimensional vector resulting from stacking the columns of $\boldsymbol{\Omega}$. 
Let $\matvec$ denote the inverse operator of $\vecmat$, that is, 
$\matvec(\vecmat(\boldsymbol{\Omega}))=\boldsymbol{\Omega}$. 
For a vector $\boldsymbol{\omega}\in\mathbb{R}^{p^2}$ with $s\leq p^2$ nonzero elements, let ${\bf D}_{{\omega}}$ denote a $p^2\times p^2$ symmetric permutation matrix putting the nonzero elements first. 
In particular, $({\bf D}_{{\omega}}\boldsymbol{\omega})_i=0$ for all $i>s$ and ${\bf D}^2={\bf I}_{p^2}$, the $p^2\times p^2$ identity matrix. 
Let $\|\boldsymbol{\omega}\|_1=\|\boldsymbol{\Omega}\|_1$ denote the usual $\ell_1$-norm and let $\|\boldsymbol{\omega}\|^{-\text{diag}}_1=\|\matvec(\boldsymbol{\omega})^{-\text{diag}}\|_1$.
Let $\otimes$ denote the Kronecker product, 
$C^\infty(\mathbb{R}^{p^2})$  the set of continuous and infinitely differentiable functions on $\mathbb{R}^{p^2}$ and let $W^{2,2}(\mathbb{R}^{p^2})$ be the Sobolev space of functions with square integrable second weak derivatives.

For any distribution $F$ on $\mathbb{R}^p$ and any \emph{scatter functional} ${\bf S}(F)$, we derive the influence function of the matrix-valued functional \begin{equation}
\bT_{ \Omega}(F) = \argmin_{\bOmega\in\mathbb{R}^{p\times p}}
\{- \text{logdet}(\boldsymbol\Omega) + \text{tr}({\bf S}(F)\boldsymbol{\Omega}) + \lambda \|\boldsymbol \Omega^{-\text{diag}}\|_1 \ \ \text{s.t.} \  \boldsymbol{\Omega}=\boldsymbol{\Omega}^\top, \boldsymbol \Omega \succ 0\}.\label{Glassomat}
\end{equation}
To do so, we rather provide the influence function of the equivalent vector-valued version 
\begin{equation}
\bT_{ \omega}(F) =
\argmin_{\boldsymbol{\omega}\in\mathbb{R}^{p^2}}
\{- \text{logdet}(\matvec(\boldsymbol{\omega})) + (\vecmat({\bf S}(F)^\top))^\top\boldsymbol{\omega} + \lambda \|\boldsymbol \omega\|^{-\text{diag}}_1 \}. \label{Glassovec}
\end{equation}

\begin{remark}
The equivalence between both definitions stems from the fact that, for any matrices ${\bf S}$ and 
$\boldsymbol{\Omega}$, $\trace({\bf S}\boldsymbol{\Omega})=(\vecmat({\bf S}^\top))^\top\vecmat(\boldsymbol{\Omega})$.
Moreover, in (\ref{Glassovec}), we do not impose  $\boldsymbol{\omega}$ to be the vectorisation of a symmetric, positive definite matrix.
The solution to the minimisation problem, without loss of generality, always respects those constraints. 
This comes from the fact that, when looking at the first order conditions, the gradient stays symmetric and that the logdet function acts as a barrier to the positive definite cone (see \citealp{sym_der}, for example). 
Therefore, it ensures that the solution (in matrix form) of the first order conditions is positive definite. 
Hence, it holds that $\bT_{{\Omega}}(F)=\matvec(\bT_{{\omega}}(F))$.
\end{remark}

\begin{remark}
The choice of the plug-in scatter functional ${\bf S}(F)$ will allow to distinguish between robust and non-robust Glasso alternatives. 
For example, the estimator provided in (\ref{Glasso}) is the empirical version of (\ref{Glassomat}) obtained by setting
$${\bf S}(F)=\mathbb{E}_F\big[({\bf X}-\mathbb{E}_F[{\bf X}])({\bf X}-\mathbb{E}_F[{\bf X}])^\top\big]\quad\textrm{and taking}\quad F=F_n.$$
Note that, unless $\lambda=0$ and $S$ is the classical covariance, there is no guarantee that $\bT_{{\Omega}}=\boldsymbol{\Omega}$ in general. $\bT_{{\Omega}}$ inherits robustness properties of the scatter functional $S$ and sparsity from the penalisation.
Other choices of robust scatter functionals will be discussed in Section~\ref{sec:IF-robust}.
\end{remark}

\begin{remark}
The expressions $\bT_{ \Omega}(F)$ and $\bT_{ \omega}(F)$ are short-hand for $\bT_{ \Omega,\lambda}(F)$ and $\bT_{ \omega,\lambda}(F)$ as they depend on the regularisation parameter $\lambda$. For simplicity, the former notation will be used and the dependence on $\lambda$ will be implicit.
\end{remark}

To derive the influence function of the Glasso, care is required
due to the non-differentiability of the penalty term.
Similarly to \cite{avella2017influence} for regression problems, we circumvent the problem by computing the influence function of the functional
\begin{equation}
\bT_{ \omega,p_m}(F) =
\argmin_{\boldsymbol{\omega}\in\mathbb{R}^{p^2}}
\{- \text{logdet}(\matvec(\boldsymbol{\omega})) + (\vecmat({\bf S}(F)^\top))^\top\boldsymbol{\omega} + \lambda p_m(\boldsymbol{\omega})\}, 
\label{Glassosmooth}
\end{equation}
where $p_m\in C^\infty(\mathbb{R}^{p^2})$ is a convex and differentiable norm which converges to $\|\cdot\|_1^{-\text{diag}}$ in $W^{2,2}(\mathbb{R}^{p^2})$ as $m\rightarrow\infty$. 
Two examples of such sequences are given by $p_m(\boldsymbol{\omega})=\|\matvec(\boldsymbol{\omega})^{-\text{diag}}\|_{k_m}$ or $p'_m(\boldsymbol{\omega})=\sum_{i\neq j}|(\matvec(\boldsymbol{\omega}))_{i,j}|^{k_m}$, with $1<k_m\in\mathbb{R}$ for all $m$ and $k_m\searrow 1$. We then define the influence function of $\bT_{\omega}$ as the limit for $m\to \infty$ of the influence function of $\bT_{ \omega,p_m}$.
The result is obtained provided the limit is independent of the chosen $p_m$ sequence, which we state in the following lemma.
\begin{lemma}
\label{lem:pm}
Let $F$ be a distribution over $\mathbb{R}^p$ and let ${\bf S}(F)$ be a scatter functional. 
For any two sequences $p_m$ and $p'_m$ converging to $\|\cdot\|_1^{-\text{diag}}$ in $W^{2,2}(\mathbb{R}^{p^2})$ and any $\bz\in\mathbb{R}^p$, it holds that
$$
\lim_{m\rightarrow\infty}\big(IF(\bz;\bT_{ \omega,p_m},F)-IF(\bz;\bT_{ \omega,p'_m},F)\big)=0.
$$
Moreover, if $\sup_{\bz\in\mathbb{R}^p}\|\vecmat(IF(\bz;S(F),F))\|<\infty$, that is, if the gross error sensitivity of $S(F)$ is bounded, then the convergence holds true uniformly:
$$
\lim_{m\rightarrow\infty}\sup_{\bz\in\mathbb{R}^p}\big(IF(\bz;\bT_{ \omega,p_m},F)-IF(\bz;\bT_{ \omega,p'_m},F)\big)=0.
$$
\end{lemma} 

The theorem below then gives the influence function of $\bT_{{\omega}}(F)$. The proof is provided in \ref{app:proofs}.
\begin{theorem}
\label{thm:IF}
Let $F$ be a distribution over $\mathbb{R}^p$ and let ${\bf S}(F)$ be a scatter functional. Let $\bz\in\mathbb{R}^p$. Assume that $\bT_{{\omega}}:=\bT_{{\omega}}(F)$ has $s$ nonzero components. Let $\bD:=\bD_{\bT_{{\omega}}}$. Then, the influence function of the graphical lasso satisfies $\bD\IF(\bz;\bT_{{\omega}},F)=$
\begin{equation*}
-\left(
\begin{array}{c}
\big(\bD(\matvec( \bT_{{\omega}})^{-1} \otimes \matvec( \bT_{{\omega}})^{-1})\bD\big)^{-1}_{1:s,1:s} (\bD\IF(\bz;\vecmat({\bf S}(F)),F))_{1:s} \\
\bf {0}
\end{array}
\right).
\end{equation*}
\end{theorem}

Theorem~\ref{thm:IF} above shows that the influence of $\bz$ on null components of $\bT_{{\omega}}$ is $0$. 
This translates to the fact that, under infinitesimal contamination, null values in the precision matrix remain zero, so that the contamination does not affect the sparsity of the solution obtained. This finding is similar to penalised regression models where the influence function is zero for the null components of the parameter vector (see, for example, \citealp{ollerer2015influence, avella2017influence}).

The influence of $\bz$ on the non-zero components of $\bT_{{\omega}}$ depends on the influence of $\bz$  on the plug-in scatter functional. A somewhat surprising conclusion at first sight is that the elements of the covariance matrix estimator corresponding to the null elements of the precision matrix do not appear in the influence function of the latter. 
This phenomenon originates from the fact that the contamination only appears through the covariance matrix estimator in the trace, more precisely, from the term
$$\trace({\bf S}(F_{\varepsilon, \bz})\boldsymbol{\Omega})=(\vecmat({\bf S}(F_{\varepsilon, \bz})^\top))^\top\boldsymbol{\omega}.$$
In turn, if $(\bT_{ \Omega}(F))_{i,j}=(\bT_{ \Omega}(F_{\varepsilon, \bz}))_{i,j}=0$, putting contamination in the cor\-responding elements of ${\bf S}(F_{\varepsilon, \bz})$ will not impact the scalar product, nor the estimation of the precision matrix. 

\subsection{Non-robustness of the Standard Glasso} \label{subsec:Glasso-nonrob}
Setting ${\bf S}(F)$ in equation \eqref{Glassovec} to be the classical covariance functional yields the standard Glasso, which we now discuss. 
First, the standard Glasso  has an unbounded influence function, hence is not robust to outliers, due to its reliance on the sample covariance. 
We illustrate its lack of robustness by computing the influence function and sensitivity curve for a setting with $p=3$ variables whose covariance matrix and corresponding sparse precision matrix are respectively given by 
\begin{equation}
    \bSigma=
    \left(
    \begin{array}{ccc}
    1 & 1/2 &1/4\\
       1/2& 1 & 1/2 \\
       1/4&1/2 &1
    \end{array}
    \right)
    \quad \text{and } \quad
    \bOmega=
      \left(
    \begin{array}{ccc}
    4/3 & -2/3 &0\\
       -2/3& 5/3 & -2/3 \\
       0&-2/3 &4/3
    \end{array}
    \right). \label{sigmatoeplitz}
\end{equation}
We introduce contamination at point $\bz= (z_1, z_2, 0)$ with values of $z_1$ and $z_2$ ranging from -6 to 6.

\begin{figure}
\centering
\includegraphics[width = 0.49\textwidth]{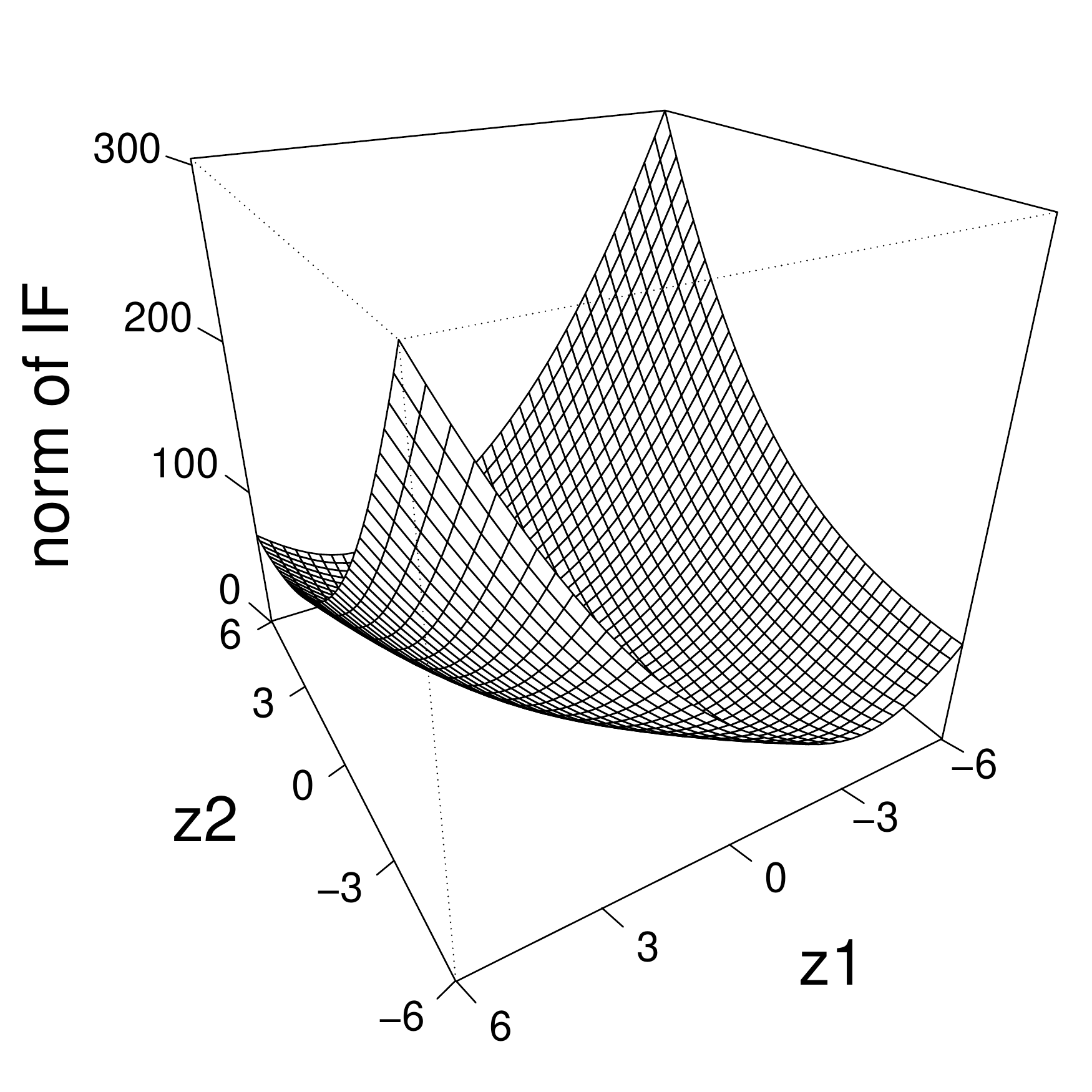}
\includegraphics[width =0.49\textwidth]{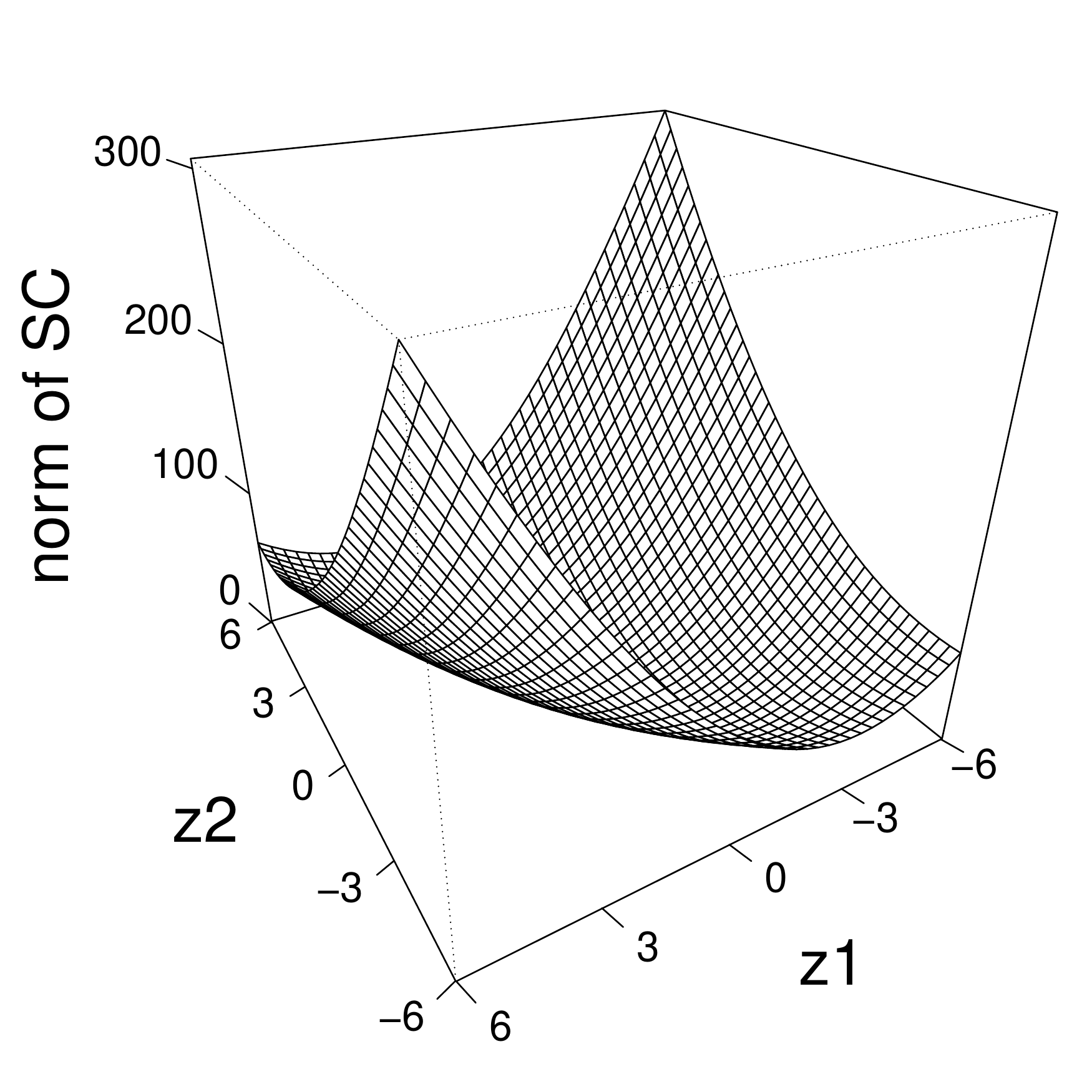}
\caption{Frobenius norm of the standard Glasso influence function (left) and sensitivity curve (right) for contamination at $(z_1, z_2, 0)$.} \label{IF-classical}
\end{figure}

The left panel of Figure \ref{IF-classical} displays $\|{\IF(\cdot;\bT_{{\Omega}},F)}\|_F$, the Frobenius norm of the influence function of the standard Glasso.
The penalty parameter is set to 
$\lambda = 8\cdot 10^{-4}$ to ensure the correct sparsity pattern is recovered.
The influence function of the standard Glasso is unbounded, hence it is not robust. This is to be expected as the influence function of the standard Glasso depends on that of the classical covariance, which is unbounded. 
The value of the norm increases fastest along the direction $z_1 = - z_2$. 

Next, we compute the sensitivity curve introduced by \cite{tukey1977EDA}. The sensitivity curve can be thought of as a finite-sample equivalent of the influence function, as in many cases the sensitivity curve converges to the influence function when the sample size goes to infinity. Suppose we have a sample $\bX_{n-1} = \{\bx_1, \ldots, \bx_{n-1}\}$ of size $n-1$ of $p$-variate observations. To this sample we add a variable point $\bz$ to obtain $\bX_z= \{\bx_1, \ldots, \bx_{n-1}, \bz\}$. We now define the sensitivity curve by
\begin{equation}
\SC{_n}({\bz}; \widehat{\bOmega};{\bX}_{n-1} ) =  n \cdot (\widehat{\bOmega}(\bX_z) - \widehat{\bOmega}(\bX_{n-1} )).
\end{equation}
It measures the change of the estimated precision matrix caused by adding the contaminated  point $\bz$ to the clean sample, standardised by $1/n$, the amount of contamination.
The right panel of Figure \ref{IF-classical} displays the Frobenius norm of the sensitivity curve for the standard Glasso (again with $\lambda = 8\cdot 10^{-4}$) for $n=1000$. The results are averaged over 50 random samples of $n-1$ clean observations drawn from a multivariate normal distribution with the covariance matrix from equation \eqref{sigmatoeplitz} and an additional contaminated data point at location $\bz = (z_1, z_2, 0)$.
The sensitivity curve of the standard Glasso is unbounded, in line with the obtained results of the influence function.

Note that the Glasso is not orthogonally equivariant for $\lambda > 0$, and this regardless of the initial covariance estimator used.
Hence, the numerical results on the Glasso influence function and sensitivity curve presented here (and in later sections for robust plug-in estimators) are dependent on the choice of covariance $\bSigma$. 
This lack of equivariance is, however, a price one needs to pay for obtaining a regularised estimate of the precision matrix.
Nonetheless, we believe that the results presented above shed light on the behavior of Glasso in the standard covariance case. Similarly, in the robust cases from Section~\ref{sec:IF-robust}, they provide valuable insights into the trade-off between robustness and statistical efficiency of the various Glasso estimators.

Next, we consider 
the gross-error sensitivity 
\begin{equation*}
    \gamma^*(\bT_{{\Omega}},F)=\underset{\bz\in\mathbb{R}^p}{\textnormal{sup}} \|{\IF(\bz;\bT_{{\Omega}},F)}\|_F,
\end{equation*}
which summarises the influence function in a single index by evaluating the maximal influence an observation may have. In the particular case of the standard Glasso, the gross-error sensitivity 
diverges since the influence function is unbounded. 
It still remains of interest, however, to determine in which direction the influence will increase the most, that is, to determine
\begin{equation*}
    \gamma^*_1(\bT_{{\Omega}},F):=\argmax_{\bz\in \mathcal{S}^{p-1}}\|{\IF(\bz;\bT_{{\Omega}},F)}\|_F,
\end{equation*}
where $\mathcal{S}^{p-1}$ denotes the hypersphere in $\mathbb{R}^p$. 
The following lemma provides the result for the unpenalised case. The solution turns out to be the eigenvector corresponding with either the smallest or the largest eigenvalue, depending on a condition on the eigenvalues themselves.
\begin{lemma}
\label{lem:maxdir}
Let $F$ be a distribution over $\mathbb{R}^p$ and let ${\bf S}(F)$ be the classical covariance functional. Fix $\lambda=0$, $g:\mathbb{R}^+\rightarrow\mathbb{R}^+:x\mapsto g(x)=x^4-2x^3$ and let $\lambda_1\geq \dotsc \geq\lambda_p$ (resp. ${\bf v}_1,\dots,{\bf v}_p$) be the eigenvalues (resp. eigenvectors) of $\bT_{{\Omega}}$. 
It holds that
\begin{equation*}
    \gamma^*_1(\bT_{{\Omega}},F)=
    \left\{
    \begin{array}{cl}
        {\bf v}_1 &\textrm{ if }g(\lambda_1)\geq g(\lambda_p)  \\
        {\bf v}_p & \textrm{otherwise}
    \end{array}
    \right.,
\end{equation*}
and
$$
\max_{\bz\in \mathcal{S}^{p-1}}\|{\IF(\bz;\bT_{{\Omega}},F)}\|_F=\sum_{i=1}^p\lambda^2_i+\max(g(\lambda_1), g(\lambda_p)).
$$
\end{lemma}

Note that, in the penalised case, it is not possible to obtain an equivalent result to Lemma~\ref{lem:maxdir} above. 
Indeed, in that situation, there is no guarantee that $\bT_{{\Omega}}=\boldsymbol{\Omega}$, so that the influence function might depend on the precision matrix in more complex forms. It is, however, easy to show that Lemma~\ref{lem:maxdir} also holds when $\lambda$ is taken large enough so that $\bT_\Omega$ is diagonal.

\section{Influence Function of Robust Glasso Estimators} \label{sec:IF-robust}
In Section \ref{subsec:boundIF} we discuss how the Glasso influence function can be bounded.
Section \ref{subsec:plugins} then presents several plug-in covariance candidates to obtain a robust Glasso and  compares them in terms of their Glasso influence function (Section \ref{subsec:IF-robust}) and sensitivity curve (Section \ref{subsec:SC-robust}).

\subsection{Bounding the Influence Function via Robust Plug-in Estimators} \label{subsec:boundIF}
In Section \ref{sec:IF-Glasso}, we show that the influence function of the Glasso depends on the influence function of the plug-in scatter functional.
To bound the influence function of a  Glasso-type of estimator, one could plug in a robust covariance estimator instead of the sample covariance in equation \eqref{Glasso}. In Section \ref{subsec:plugins}, we discuss several candidates for robust covariance plug-ins.
Such robust Glasso estimators would indeed result in a bounded influence function since the gross-error sensitivity can be bounded as
$$
    \gamma^*(\bT_{{\omega}},F)\leq \left\|(\bD(\matvec( \bT_{{\omega}})^{-1} \otimes \matvec( \bT_{{\omega}})^{-1})\bD\big)^{-1}_{1:s,1:s}\right\|\gamma^*((\bD\vecmat({\bf S}))_{1:s},F),
$$
where the norm used is the operator norm. 
Note that, should \mbox{$(\bT_{{\omega}}(F))_{i,j}\neq 0$} everywhere (implying $\bD =\mathbf{I}_p$), this bound simplifies to
$$
\gamma^*(\bT_{{\omega}},F)\leq \|\bT_{\Omega}\|^2\gamma^*((\vecmat({\bf S}))_{1:s},F).
$$

\subsection{Robust Covariance Plug-ins} \label{subsec:plugins}
As candidates for the robust covariance estimator to be plugged into the Glasso objective function, we consider the popular, traditional row-wise robust Minimum Covariance Determinant (MCD) estimator 
\citep{rousseeuw1984least, rousseeuw1985multivariate, rousseeuw1999fast} as well as a robust covariance estimator based on pairwise robust correlation estimators (see e.g.  \citealp{ollerer2015robust, croux2016robust}). Unlike the MCD, the latter are frequently used in the literature on Glasso robustifications as they are available in high-dimensional settings and protect against cellwise outliers \citep{alqallaf2009propagation}. 

The influence function of the Glasso based on the MCD\footnote{Note that, throughout this paper, MCD is defined using $75\%$ of the data, so as to achieve a balance between robustness and efficiency. For the sake of comparison, we also considered the reweighted MCD, with the same $25\%$ breakdown point.} can simply be obtained by plugging in the expression of the MCD influence function (see \citealp{croux1999influence}) into the expression of the Glasso influence function provided in Theorem \ref{thm:IF}. 
To arrive at the Glasso influence function based on a pairwise robust correlation estimator, we need to introduce additional notation related to the latter's functional representation. 

Denote with $S$ a scale functional and $R$ a correlation functional. The definition of the covariance matrix functional ${\bf T}_{\Sigma}$ based on pairwise robust correlation estimators is given by 
\begin{equation}\label{pairwise_cov}
    ({\bf T}_{\Sigma}(F))_{j,k} = S(F_j) S(F_k) R(F_{j,k}),
\end{equation}
where $F_j$ and $F_k$ denote the marginal distributions of the $j$th and $k$th components respectively, and $F_{j,k}$ denotes the bivariate (marginal) distribution of the $j$th and $k$th components.

As a scale estimator, we use the $Q_n$ estimator \citep{rousseeuw1993alternatives}, whose functional version is given by $S(F) = cH^{-1}\left(\frac{1}{4}\right)$, where $c$ is a constant ensuring Fisher-consistency and $H$ is the distribution function of the absolute difference of two independent random variables, each with the same distribution $F$.

As robust correlation estimator, we consider Spearman's rank correlation \citep{spearman1904}, Kendall's correlation coefficient \citep{kendall1938new}, the Gaussian rank correlation \citep{boudt2012gaussian} and the Quadrant correlation \citep{blomqvist1950measure}. As our main focus is the influence function, we will consider the functional version of these estimators below. For the sake of completeness, we give the more common finite-sample definitions of these estimators in \ref{app:robcor}.

The functional form of Spearman's rank correlation coefficient $R_S$ is given by
\begin{equation*}
     R_S(F_{j,k}) = 12\E_{F_{j,k}}\left[F_{j}\left(X_j\right)F_{k}\left(X_k\right)\right] - 3,
\end{equation*}
where $(X_j, X_k) \sim F_{j, k}$ and $F_j$ and $F_k$ are the marginal distributions of $X_j$ and $X_k$ as before. Kendall's correlation coefficient $R_K$ is given in functional form by
\begin{equation*}
     R_K(F_{j,k}) = \E_{F_{j,k}}\left[\sign((X_{j_1} - X_{j_2})(X_{k_1} - X_{k_2}))\right],
\end{equation*}
where  $(X_{j_1}, X_{k_1})$ and $(X_{j_2}, X_{k_2})$ are independent bivariate random variables, each with distribution $F_{j,k}$. The Gaussian rank correlation coefficient is given in functional form by
\begin{equation*}
     R_G(F_{j,k}) = \E_{F_{j,k}}\left[\Phi^{-1}\left(F_{j}\left(X_j\right)\right) \Phi^{-1}\left(F_{k}\left(X_k\right)\right)\right],
\end{equation*}
where $\Phi^{-1}$ denotes the quantile function of the standard normal distribution. Note that if $F_{j, k}$ is bivariate normal, we obtain the functional of the Pearson correlation coefficient. The Quadrant correlation coefficient is given in functional form by
\begin{equation*}
     R_Q(F_{j,k}) = \E_{F_{j,k}}\left[\sign((X_j - \median(X_j))(X_k - \median(X_k)))\right].
\end{equation*}

For comparability of the influence functions of the Glasso based on the different plug-in covariances, it is important that all estimate the same population quantity, i.e.\ are Fisher consistent. 
Hence, we  use the Fisher consistent versions of the Spearman, Kendall and Quadrant correlation at the normal model by applying the non-linear transformations as given, for instance, in \cite{croux2010influence}. 
In particular, we use 
$\tilde{R}_S= 2\sin\left(\frac{\pi}{6}R_S\right)$, $\tilde{R}_K = \sin\left(\frac{\pi}{2}R_K\right)$ and $\tilde{R}_Q = \sin\left(\frac{\pi}{2}R_Q\right)$.
No transformation for the Gaussian rank correlation is needed as it is Fisher consistent at the normal model. Note that in finite samples, these transformations can destroy the positive semidefiniteness (PSD) of the resulting covariance matrix. 
This is typically dealt with by projecting the estimate onto the space of PSD matrices, for example with respect to the Frobenius norm (as in \citealp{ollerer2015robust}) or the elementwise $\ell_{\infty}$ norm (as in \citealp{loh2018high}). 
This does not affect the discussion on the influence function.

The influence function of the resulting covariance estimator is then given by
\begin{align*}
 \IF(\bz;(\bT_{\Sigma}(F))_{j,k},F) =& \IF(z_j;S,F_j)S(F_k)R(F_{j,k}) \\
    +& S(F_j)\IF(z_k;S,F_k)R(F_{j,k})\\
    +&S(F_j)S(F_k)\IF(\bz;R,F_{j,k}).
\end{align*}
To compute this influence function, the expression of the influence function of the scale estimator, namely the $Q_n$, and the considered correlation estimators are thus needed.
The former is available in \cite{rousseeuw1993alternatives}, the latter in \cite{croux2010influence} for Kendall, Spearman and  Quadrant, and \cite{boudt2012gaussian} for the Gaussian rank.

In the next subsections, we present the  influence functions and sensitivity curves of the Glasso estimators based on these different robust covariances.

\subsection{Influence Functions} \label{subsec:IF-robust}
We consider the same example as in Section \ref{subsec:Glasso-nonrob} and obtain the influence function of the Glasso (with the same fixed penalty parameter as before)
computed from five different robust covariance plug-ins, namely the MCD, and the pairwise correlation estimators based on Spearman, Kendall, Gaussian rank and Quadrant.

\begin{figure}[!t]
  \begin{minipage}[b]{0.32\linewidth}
    \centering
    \includegraphics[width=\linewidth]{S_3D_IF.pdf} 
    \caption*{Classical}
  \end{minipage}
  \begin{minipage}[b]{0.32\linewidth}
    \centering
    \includegraphics[width=\linewidth]{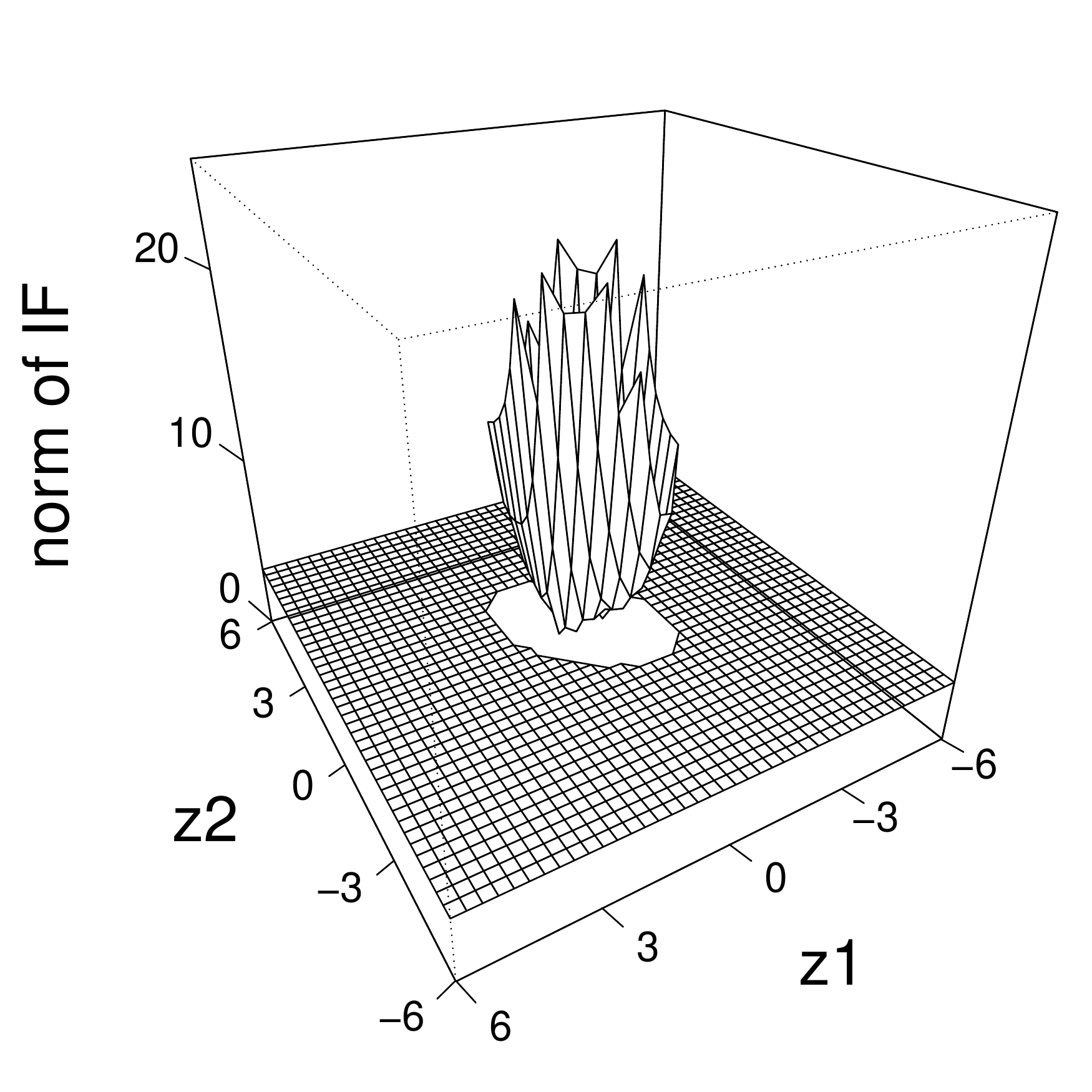} 
    \caption*{MCD}
   \end{minipage} 
  \begin{minipage}[b]{0.32\linewidth}
    \centering
    \includegraphics[width=\linewidth]{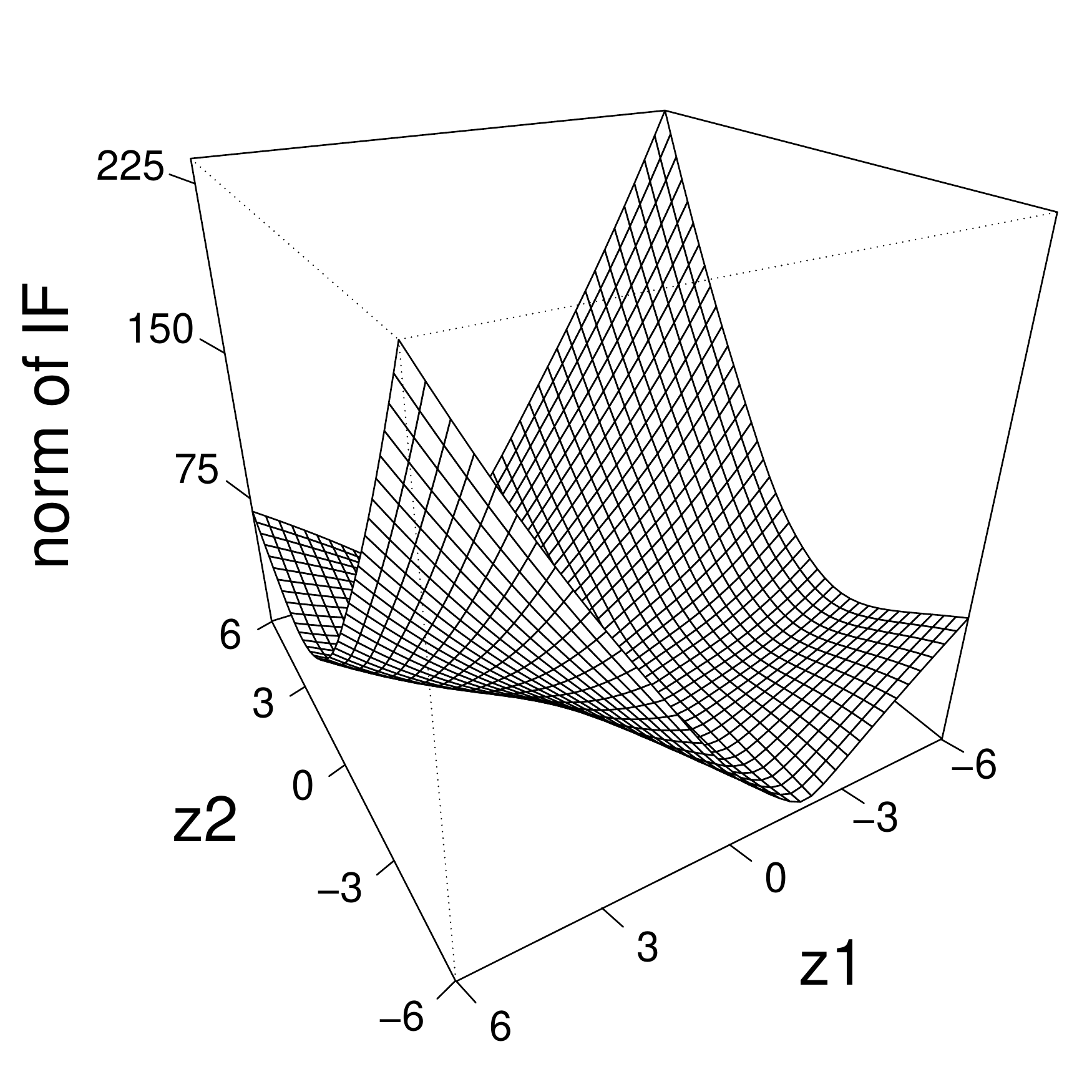} 
    \caption*{Gaussian rank}
   \end{minipage}   
  
   \begin{minipage}[b]{0.32\linewidth}
    \centering
    \includegraphics[width=\linewidth]{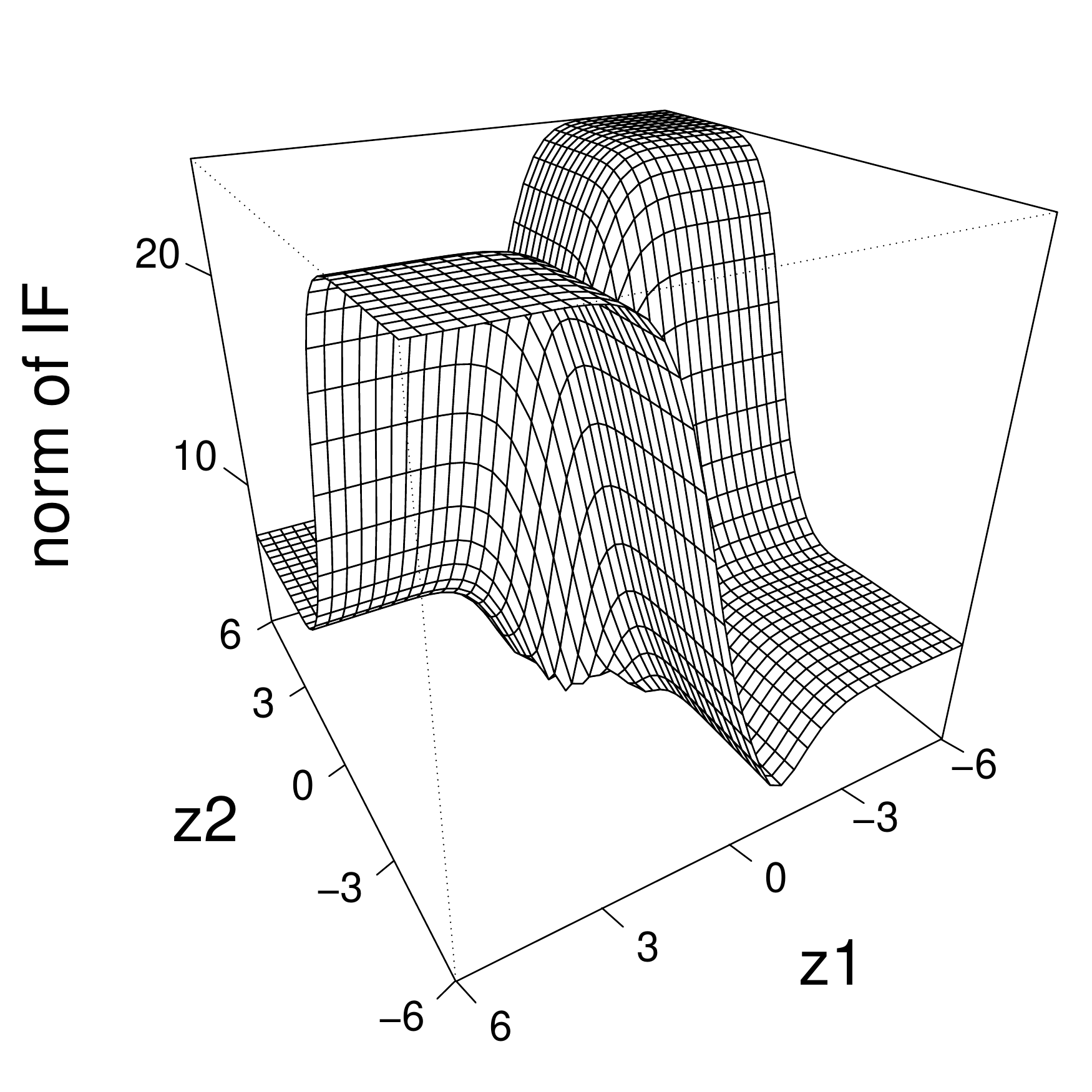} 
    \caption*{Kendall}
  \end{minipage}
     \begin{minipage}[b]{0.32\linewidth}
    \centering
    \includegraphics[width=\linewidth]{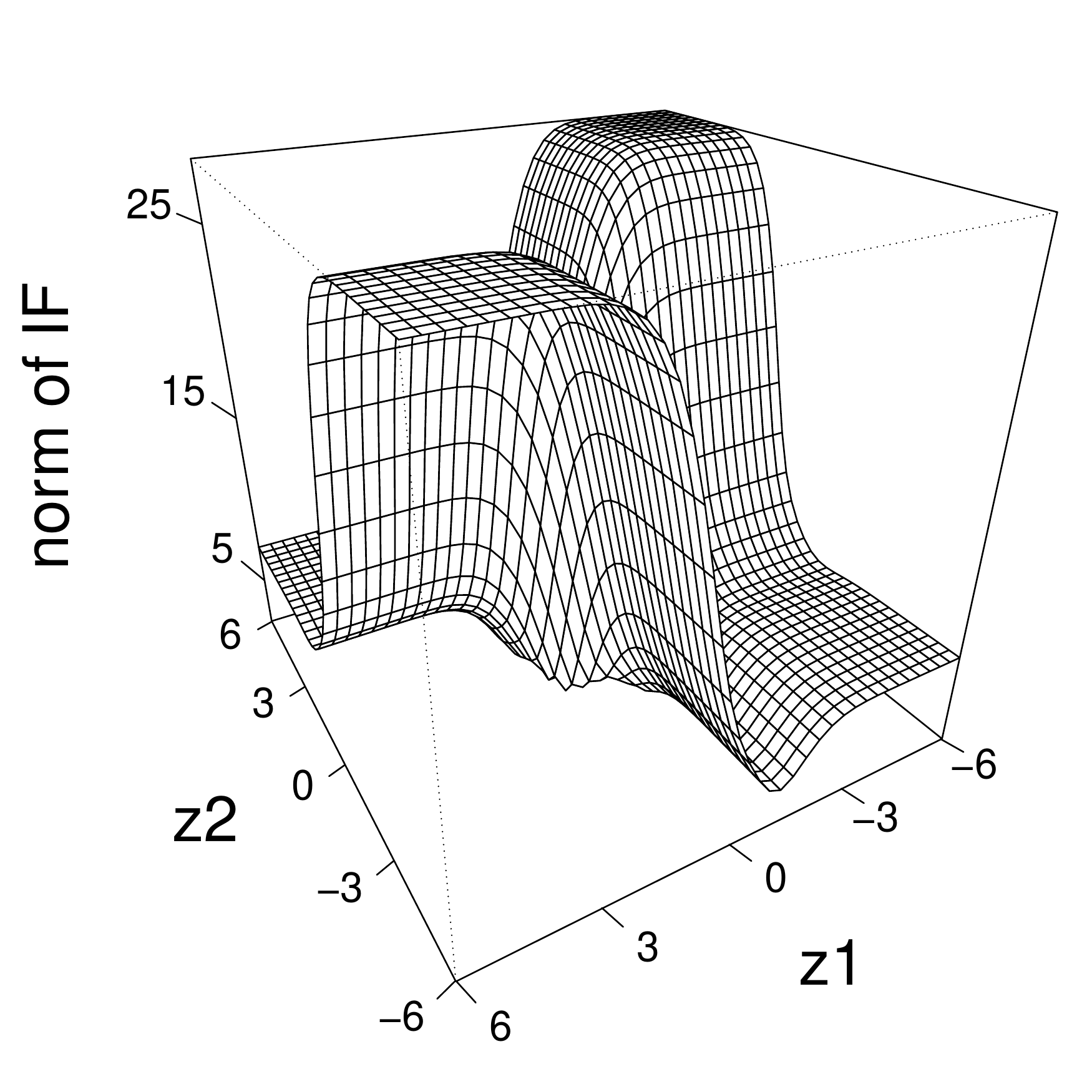} 
    \caption*{Spearman}
  \end{minipage}
  \begin{minipage}[b]{0.32\linewidth}
    \centering
    \includegraphics[width=\linewidth]{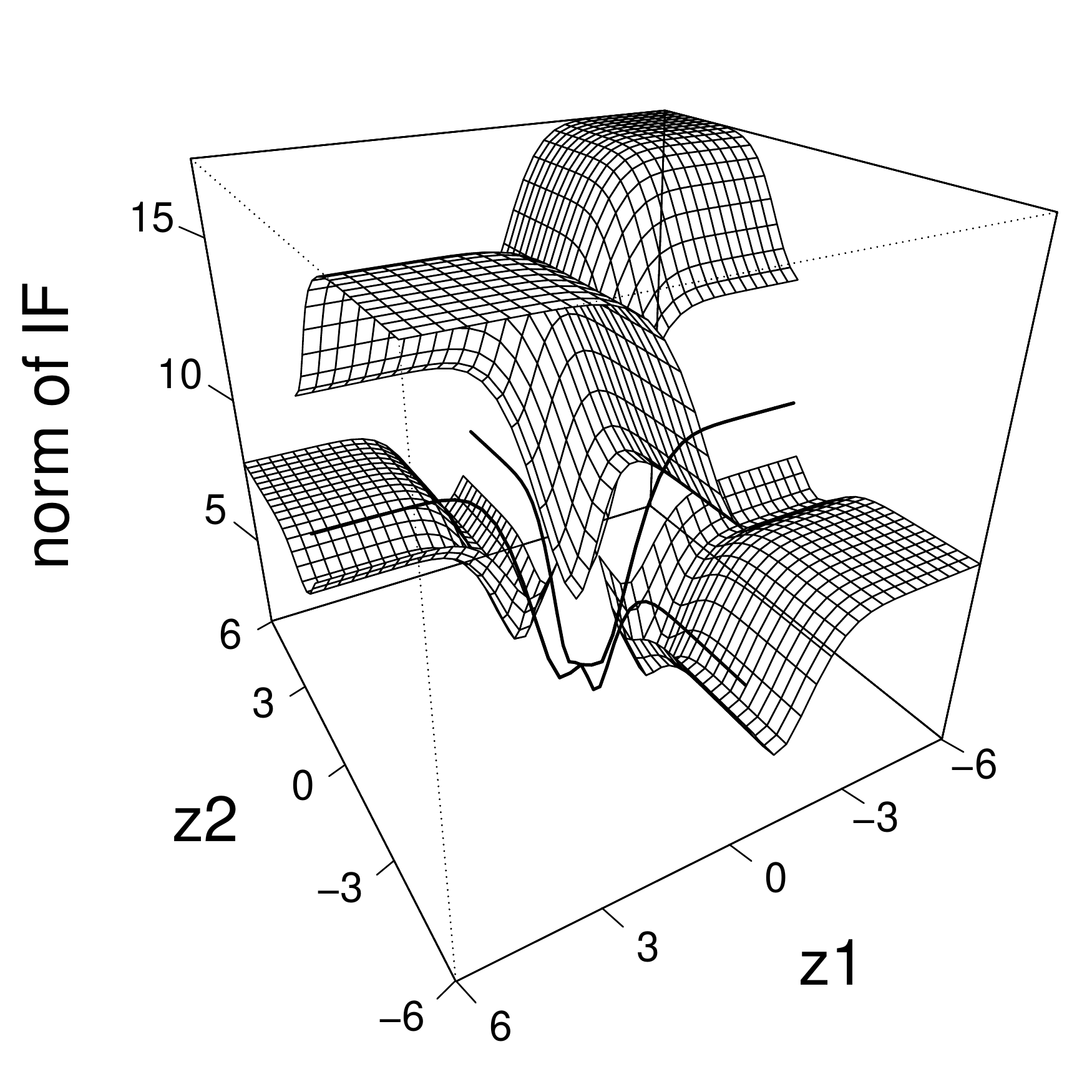} 
    \caption*{Quadrant}
  \end{minipage} 
  \caption{Norm of the Glasso influence function based on several covariance matrix estimators for contamination at $(z_1, z_2, 0)$.} \label{3D_IFnorm}
  \label{3DIF}
\end{figure}

Figure \ref{3D_IFnorm} displays the norm of the Glasso influence functions for the various covariance estimators, and includes the one for the classical covariance (from Figure \ref{IF-classical}) to facilitate comparisons.
When using the MCD as initial estimator, we see that the norm of the influence function resembles the one for the standard Glasso around the center, while the influence becomes bounded for points further away from it. 
The Glasso influence function based on the MCD as initial estimator is thus B-robust (see \citealp{hampel86}).
However, while being highly robust, it should be noted that it is computationally demanding to obtain and, in practice, only available when $p$ is small relative to $n$. Note that the same comments apply for the reweighted version of MCD, see Figure~\ref{IF_RMCD} of \ref{app:figs_SC}.

Next, we turn to the Glasso influence function based on the Gaussian rank. 
At first sight, it largely resembles the results for the classical covariance, as its influence function is also unbounded.
This is not surprising, as the influence function of the Gaussian rank correlation is unbounded \citep{boudt2012gaussian}.
Nonetheless, the values of the norm are smaller, thereby indicating a somewhat larger resistance to single outliers than the classical covariance. 
This is most likely due to the use of $Q_n$ as an estimator of the marginal scales. 
This unboundedness of the influence function is somewhat misleading however, and  our results on the sensitivity curve (to be discussed in Section \ref{subsec:SC-robust}) indicate that the Gaussian rank yields a substantially more robust estimator than the standard Glasso in finite samples.

The Glasso influence functions based on Kendall and Spearman  are very similar: they are bounded and smooth, hence the Glasso computed from these initial covariance matrices is robust to outliers. 
Finally, the Glasso influence function based on the Quadrant is also bounded but not smooth: it has jumps at the coordinate axes due to the occurrence of the sign function and the median in its definition. 
Hence, small changes in data points close to the median of one of the marginals may lead to relatively large changes in its Glasso.
Nonetheless, usage of the Quadrant as initial estimator results in the overall smallest values of the norm of the Glasso influence function.
This is in line with findings in the setting of correlation estimation where the Quadrant is known to be the ``most B-robust", i.e.\ it has  the lowest gross-error sensitivity among a class of correlation estimators based on product moments (hence not including the MCD) \citep{raymaekers2021fast}.

\begin{figure}[t]
  \begin{minipage}[b]{0.32\linewidth}
    \centering
    \includegraphics[width=\linewidth]{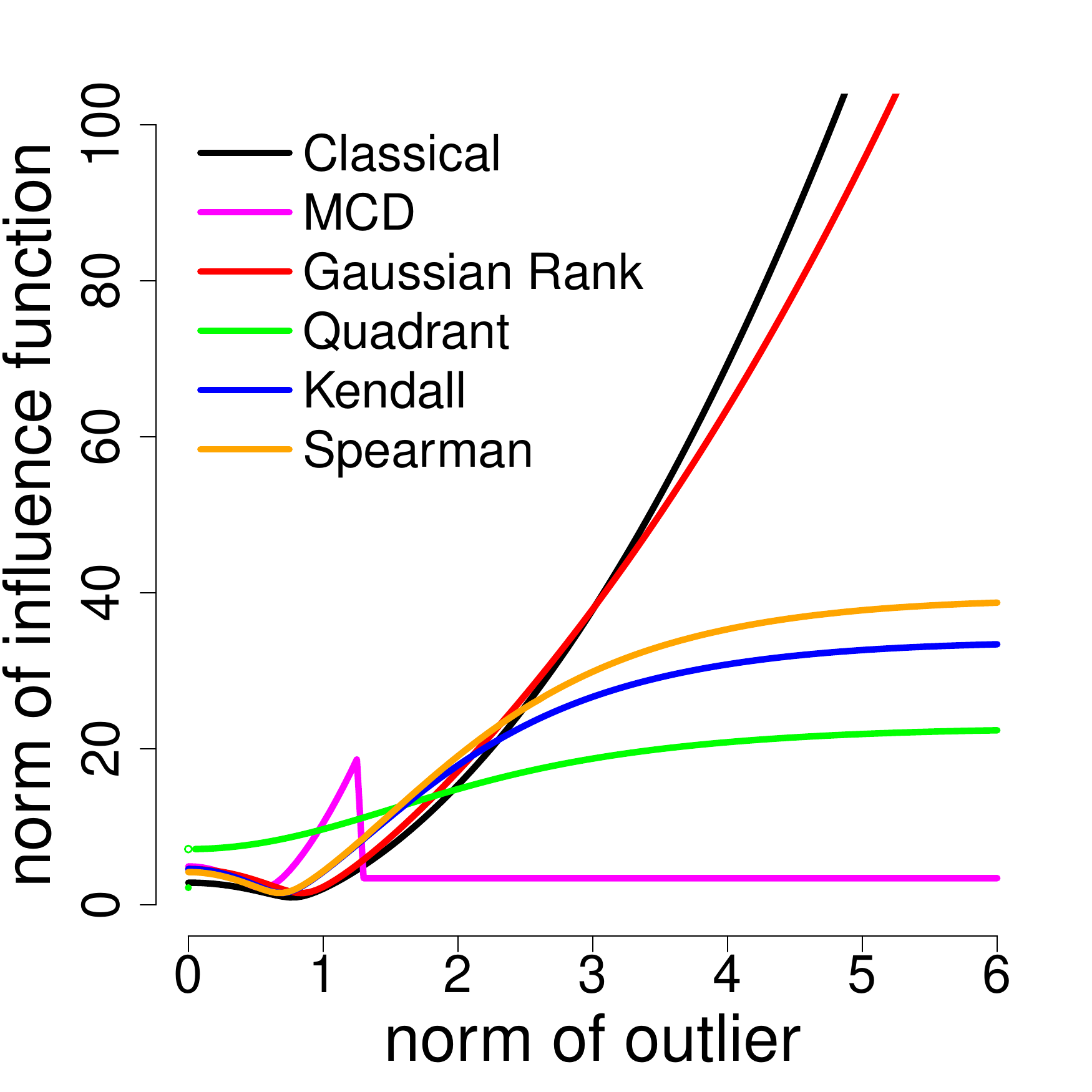} 
  \end{minipage}
  \begin{minipage}[b]{0.32\linewidth}
    \centering
    \includegraphics[width=\linewidth]{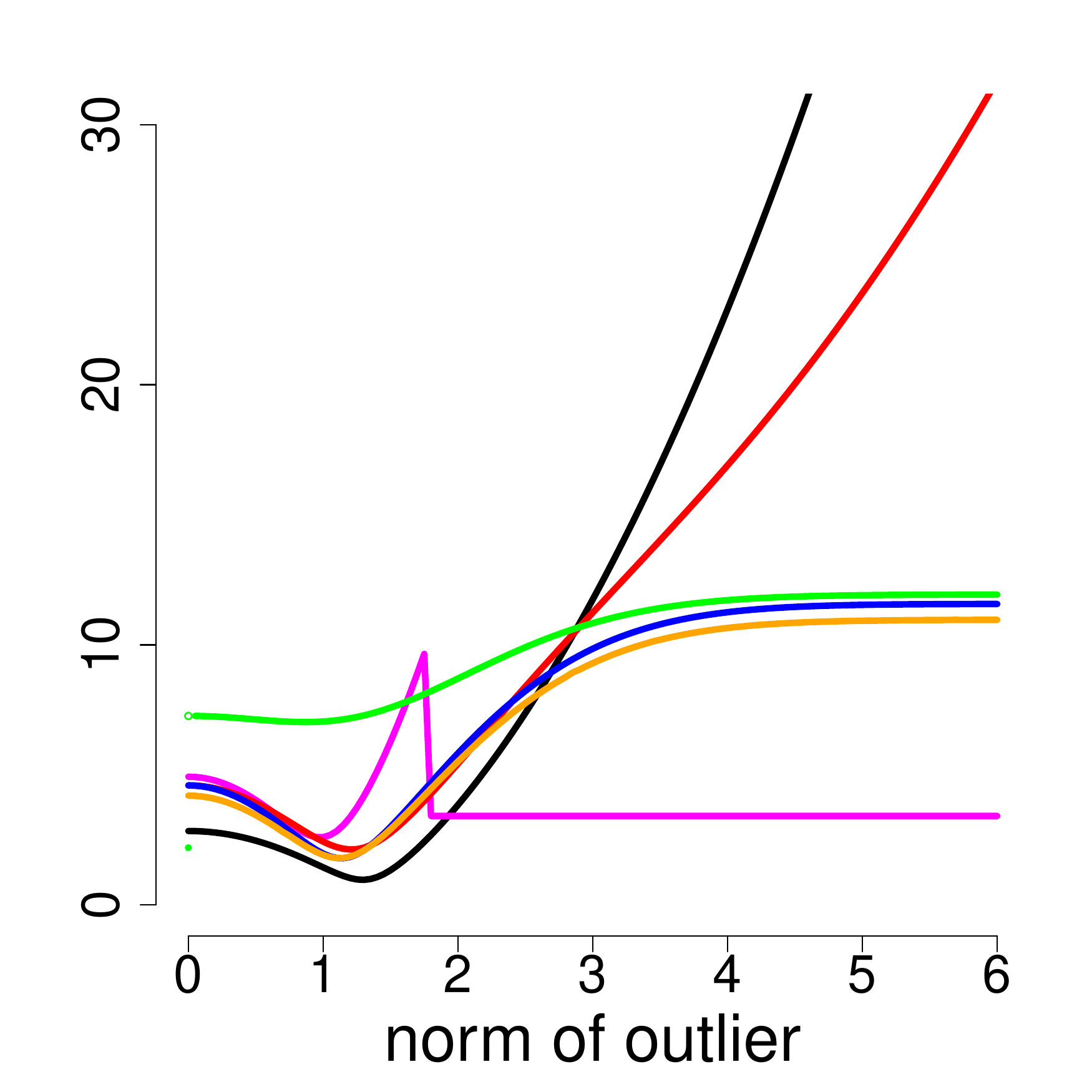} 
   \end{minipage} 
  \begin{minipage}[b]{0.32\linewidth}
    \centering
    \includegraphics[width=\linewidth]{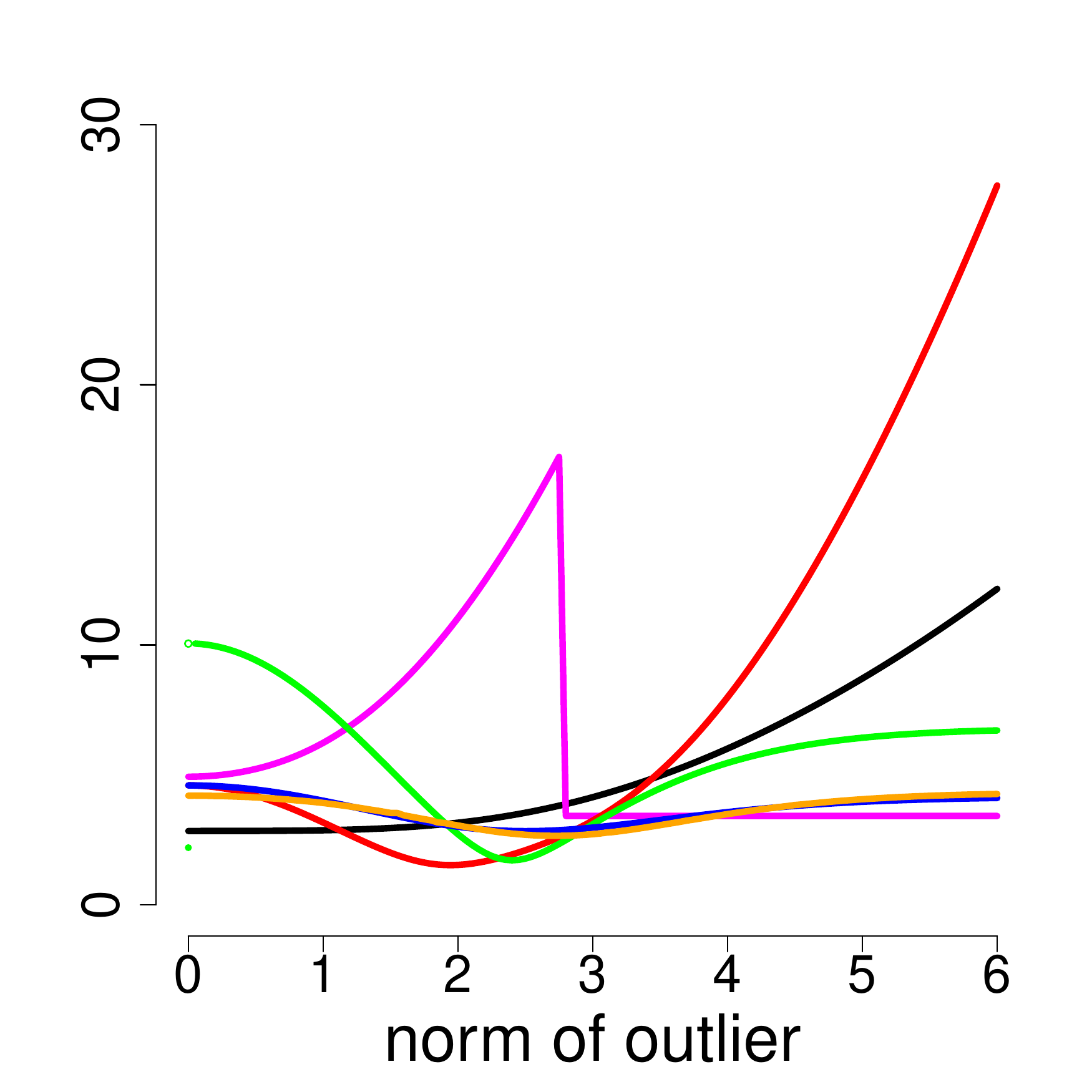} 
   \end{minipage}   
   \caption{Norm of the Glasso (with $\lambda = 8\cdot 10^{-4}$) influence function based on several covariance matrix estimators for contamination in the direction of the eigenvector corresponding to the first (left), second (middle) and third (right) eigenvalue of $T_\Omega$ based on the true covariance matrix.}
   \label{2D_norm_IF}
\end{figure}

We summarise the performance of the Glasso estimators based on the different initial covariances in Figure \ref{2D_norm_IF}. 
Figure \ref{2D_norm_IF} displays the norm of the Glasso influence functions for increasing values of the outlier's norm where the contamination is placed in the direction of the eigenvectors of $\bT_\Omega$ based on the true covariance matrix. 
In line with our previous discussion, we see that the influence function for the standard Glasso  and  the Glasso based on the Gaussian rank are unbounded. The Glasso influence function based on the MCD, Quadrant, Spearman and Kendall are all bounded, with the former two being most robust for more severe outliers. Finally, much like the unpenalised case described in Lemma~\ref{lem:maxdir}, the influence seems higher in the direction associated with the largest eigenvalue of $\bT_\Omega$ (left panel Figure \ref{2D_norm_IF}).

\subsection{Sensitivity Curves} \label{subsec:SC-robust}
We compute the Glasso sensitivity curves based on pairwise correlation estimators computed from the Gaussian rank, Spearman and Kendall correlation, analogously to the procedure described in Section \ref{subsec:Glasso-nonrob} for the sample covariance\footnote{We omit results for the MCD as it suffers from approximation errors when using the classical FastMCD algorithm and is computationally demanding to compute exactly. Likewise, the Quadrant case is omitted as its discontinuity leads to unstable behaviour in finite samples.}.
Note that, contrary to the influence functions which are always well-defined, the sensitivity curves can only be computed provided $S(F_n)$ exists. In particular, this requires $p<n$ in some cases (classical covariance function, MCD).
Results are available in Figures \ref{2D_norm_SC} and  \ref{3DSC}  of \ref{app:figs_SC}.

The Glasso sensitivity curves for the Spearman and Kendall correlation are bounded and alike, in line with the results of Section \ref{subsec:IF-robust}. The Glasso computed from the Gaussian rank is considerably more robust than the standard Glasso in finite samples. Nonetheless, its influence function does not reflect this as it is unbounded (see Figure \ref{3DIF}).
This apparent contradiction was discussed by \cite{boudt2012gaussian} for the Gaussian rank correlation estimator. More precisely, for any fixed sample size, the sensitivity curves for the Gaussian rank correlation and the Pearson correlation are both bounded due to the boundedness of the correlation coefficient itself. As the sample size increases, however, the finite-sample gross-error sensitivities of both estimators diverge to infinity which corresponds with their unbounded influence functions. The key difference is that the Pearson correlation has a finite-sample gross-error sensitivity diverging at a rate of $\sim n$, whereas that of the Gaussian rank correlation diverges much slower at a rate of $\sim \log(n)$. This explains the difference in robustness, which translates to the Glasso setting.
Still, the Glasso computed from the Spearman and Kendall correlations are considerably more robust.

\section{Asymptotic Variances} \label{sec:ASV}
Apart from robustness, it is desirable that an estimator has a high statistical efficiency. 
To measure the statistical efficiency of the different Glasso estimators, we first provide the expressions of their asymptotic variances, then compare the asymptotic variances to the standard Glasso.

Should a Bahadur representation of $\bT_{\Omega}(F)$ exist, one would then conclude to asymptotic normality and Fr\'echet differentiability of the precision matrix estimator. 
The asymptotic variances would then be given by
$$
\text{ASV}((\bT_\Omega)_{i,j},F)=\int_{\mathbb{R}^p}\text{IF}^2(\bz;(\bT_{\Omega})_{i,j},F) dF(\bz),
$$
for any $(i,j)$ such that $(\bT_\Omega(F))_{i,j}\neq 0$. 
In matrix form, plugging in the expression of the influence function derived in Theorem~\ref{thm:IF} yields
\begin{equation}
\label{eq:ASVmat}
\text{ASV}((\bD\bT_\omega)_{1:s},F)=\boldsymbol{A} (\bD\text{ASV}(\vecmat({\bf S}(F)),F)\bD)_{1:s}\boldsymbol{A}^\top,
\end{equation}
with 
$$
\boldsymbol{A}=\big(\bD(\matvec( \bT_{{\omega}})^{-1} \otimes \matvec( \bT_{{\omega}})^{-1})\bD\big)^{-1}_{1:s,1:s}.
$$
A formal verification of the Bahadur representation seems to be challenging and is still an open question. For example, in the context of scatter estimation, results include \cite{heshao1996} (for $M$-estimation) or  \cite{catorlopuhaa2010} (for the MCD estimator). Asymptotic normality results in the graphical models context are typically based on minimax bounds (see, for example, \cite{Renasymptotic} and references therein).

We now turn to the numerical example introduced in Section \ref{subsec:Glasso-nonrob}, and compute the asymptotic variances for the Glasso based on the MCD, and pairwise correlation estimators. 
Table~\ref{ASV-example} summarises the asymptotic efficiencies relative to the standard Glasso  for three different components--- diagonal elements (1,1), (2,2) and off-diagonal element (2,1) ---of the precision matrix. 
Results on the other components are similar and therefore omitted. The values from Table~\ref{ASV-example} were obtained from (\ref{eq:ASVmat}) using numerical integration.
The asymptotic variances obtained above are consistent with simu\-lation results conducted by the authors. 

The Glasso based on Gaussian rank returns, overall, the highest efficiencies but is closely followed by the Glasso based on the Kendall and Spearman correlations. 
Their asymptotic efficiencies remain above 80\%. 
We suspect that the efficiency of around 80\% for the diagonal elements stems mainly from the fact that we are using the $Q_n$ estimator to estimate the diagonal elements of our initial covariance matrix. 
The $Q_n$ has an efficiency of 82\% at Gaussian data  \citep{rousseeuw1993alternatives}. The Glasso based on Quadrant correlation and MCD, on the other hand, suffer from a severe loss of statistical efficiencies with drops to around 30\%-40\% for the former and to around 25-30\% for the MCD. The reweighted MCD displays considerably higher efficiencies of around $70\%$.

\begin{table}
\centering
\caption{Asymptotic efficiencies of the Glasso estimator based on several robust covariance matrix estimators relative to the standard Glasso and this for three components of the precision matrix.}
\begin{tabular}{cccccccc} 
 \hline
&& \multicolumn{6}{c}{Glasso with covariance estimator} \\
Component && Gaussian  & Kendall & Spearman & Quadrant & MCD & Reweighted \\ 
&&  rank &  &  &  &  & MCD \\
 \hline
  (1,1)&& 0.8210 & 0.8150 & 0.8097 & 0.4866 & 0.3003 & 0.6755\\ 

 (2,2)&& 0.8085 & 0.8091 & 0.8027 & 0.4187 & 0.3076 & 0.6709\\

  (2,1) && 0.9563 & 0.8725 & 0.8491 & 0.3004 &  0.2556  & 0.7101\\
  \hline
\end{tabular} 
\label{ASV-example}
\end{table}

\section{Conclusion} \label{sec:conclusion}
The standard Glasso is one of the most often used sparse estimators for precision matrices yet it may seriously be affected by the presence of even a single outlier.
In this paper, we study the robustness of the Glasso by deriving expressions of its influence function for any plug-in scatter functional.
We show that the influence function of the standard Glasso is unbounded, thereby proving its lack of robustness.
Nonetheless, the Glasso influence function can be easily bounded by plugging in a robust covariance matrix estimate instead of the sample covariance.

We consider several robust Glasso estimators relying on different  covariance matrix plug-ins (namely the MCD or pairwise correlation estimators based on Spearman, Kendall, Gaussian rank or Quadrant), and study their trade-off in robustness versus statistical efficiency.
When plugging in a pairwise robust correlation  estimate based on Kendall or Spearman, the resulting 
Glasso combines the attractive properties of
(i) a bounded and smooth influence function, and
(ii) high statistical efficiency at the normal model.
Opting for the Glasso based on the Gaussian rank leads to a higher statistical efficiency, better 
protection against single outliers than the standard Glasso in finite samples but a price is paid in terms of robustness compared to the Glasso  based on Spearman or Kendall.
Opting for the Glasso based on the MCD or Quadrant correlation, on the other hand, results 
in a very strong resistance against outliers but low statistical efficiency though the efficiency of the Glasso MCD is considerably increased by considering the reweighted MCD.
Besides, the MCD is computationally demanding and only available for $p<n$, thereby limiting its applicability to low-dimensional settings.

An appropriate choice of initial covariance matrix estimate to be plugged-into the Glasso thus depends on
the trade-off between robustness and statistical efficiency the practitioner is willing to make. As general advice to applied statisticians, we recommend to first apply the standard Glasso as well as a Glasso estimator with a highly-robust covariance plug-in. 
If both yield very different results, then outliers are likely present and one could opt for a Glasso estimator with strong resistance against outliers.
If both yield similar results, then one could opt for a robust Glasso estimator with high statistical efficiency.

The presented results on the influence function of Glasso allow for the study of (optimal) B-robustness within classes of estimators \citep[p.\ 116]{hampel86}. For example, a relevant question would be whether we can find the estimator with the highest efficiency given a bound on the gross-error sensitivity, within the class of Glasso estimators based on pairwise initial covariance estimators. In addition to B-robustness, also V-robustness \citep[p.\ 128]{hampel86} could be studied. We consider these interesting directions for future research.

\bigskip
\noindent
\textbf{Acknowledgements.}
We acknowledge support through the HiTEc Cost Action CA21163.
The first and third author's work is supported by the Action de Recherche Concertée ARC-IMAL.
The last author is supported by the Dutch Research Council (NWO) under grant number VI.Vidi.211.032. 

\bibliographystyle{asa}
\bibliography{mybibfile} 

\begin{thebibliography}{34}
\newcommand{\enquote}[1]{``#1''}
\expandafter\ifx\csname natexlab\endcsname\relax\def\natexlab#1{#1}\fi

\bibitem[{Aerts and Wilms(2017)}]{aerts2017cellwise}
Aerts, S. and Wilms, I. (2017), \enquote{Cellwise robust regularized
  discriminant analysis,} \textit{Statistical Analysis and Data Mining: The ASA
  Data Science Journal}, 10, 436--447.

\bibitem[{Alqallaf et~al.(2009)Alqallaf, Van~Aelst, Yohai, and
  Zamar}]{alqallaf2009propagation}
Alqallaf, F.; Van~Aelst, S.; Yohai, V.~J. and Zamar, R.~H. (2009),
  \enquote{Propagation of outliers in multivariate data,} \textit{The Annals of
  Statistics}, 37, 311--331.

\bibitem[{Avella-Medina(2016)}]{avellathesis}
Avella-Medina, M. (2016), \enquote{Robust penalized M-estimators for
  generalized linear and additive models,} Ph.D. thesis, Geneva University.

\bibitem[{Avella-Medina(2017)}]{avella2017influence}
--- (2017), \enquote{Influence functions for penalized ${M}$-estimators,}
  \textit{Bernoulli}, 23, 3178--3196.

\bibitem[{Banerjee et~al.(2008)Banerjee, El~Ghaoui, and
  d'Aspremont}]{banerjee2008model}
Banerjee, O.; El~Ghaoui, L. and d'Aspremont, A. (2008), \enquote{Model
  selection through sparse maximum likelihood estimation for multivariate
  {G}aussian or binary data,} \textit{Journal of Machine Learning Research}, 9,
  485--516.

\bibitem[{Blomqvist(1950)}]{blomqvist1950measure}
Blomqvist, N. (1950), \enquote{On a measure of dependence between two random
  variables,} \textit{The Annals of Mathematical Statistics}, 21, 593--600.

\bibitem[{Boudt et~al.(2012)Boudt, Cornelissen, and Croux}]{boudt2012gaussian}
Boudt, K.; Cornelissen, J. and Croux, C. (2012), \enquote{The {G}aussian rank
  correlation estimator: {R}obustness properties,} \textit{Statistics and
  Computing}, 22, 471--483.

\bibitem[{Cator and Lopuha\"a(2010)}]{catorlopuhaa2010}
Cator, E. and Lopuha\"a, H. (2010), \enquote{Asymptotic expansion of the
  minimum covariance determinant estimator,} \textit{J. Mult. Anal.}, 101,
  2372--2388.

\bibitem[{Croux and Dehon(2010)}]{croux2010influence}
Croux, C. and Dehon, C. (2010), \enquote{Influence functions of the {S}pearman
  and {K}endall correlation measures,} \textit{Statistical methods \&
  applications}, 19, 497--515.

\bibitem[{Croux and Haesbroeck(1999)}]{croux1999influence}
Croux, C. and Haesbroeck, G. (1999), \enquote{Influence function and efficiency
  of the minimum covariance determinant scatter matrix estimator,}
  \textit{Journal of Multivariate Analysis}, 71, 161--190.

\bibitem[{Croux and {\"O}llerer(2016)}]{croux2016robust}
Croux, C. and {\"O}llerer, V. (2016), \enquote{Robust and sparse estimation of
  the inverse covariance matrix using rank correlation measures,} in
  \textit{Recent Advances in Robust Statistics: Theory and Applications},
  Springer, pp. 35--55.

\bibitem[{Fan et~al.(2014)Fan, Han, and Liu}]{fan2014challenges}
Fan, J.; Han, F. and Liu, H. (2014), \enquote{Challenges of big data analysis,}
  \textit{National Science Review}, 1, 293--314.

\bibitem[{Finegold and Drton(2011)}]{finegold2011robust}
Finegold, M. and Drton, M. (2011), \enquote{Robust graphical modeling of gene
  networks using classical and alternative $t$-distributions,} \textit{The
  Annals of Applied Statistics}, 5, 1057--1080.

\bibitem[{Friedman et~al.(2008)Friedman, Hastie, and
  Tibshirani}]{friedman2008sparse}
Friedman, J.; Hastie, T. and Tibshirani, R. (2008), \enquote{Sparse inverse
  covariance estimation with the graphical lasso,} \textit{Biostatistics}, 9,
  432--441.

\bibitem[{Hampel et~al.(1986)Hampel, Ronchetti, Rousseeuw, and
  Stahel}]{hampel86}
Hampel, F.~R.; Ronchetti, E.~M.; Rousseeuw, P.~J. and Stahel, W.~A. (1986),
  \textit{Robust statistics: the approach based on influence functions}, John
  Wiley \& Sons.

\bibitem[{He and Shao(1996)}]{heshao1996}
He, X. and Shao, Q.-M. (1996), \enquote{A general Bahadur representation of
  M-estimators and its application to linear regression with nonstochastic
  designs,} \textit{Ann. Stat.}, 24, 2608--2630.

\bibitem[{Kendall(1938)}]{kendall1938new}
Kendall, M.~G. (1938), \enquote{A new measure of rank correlation,}
  \textit{Biometrika}, 30, 81--93.

\bibitem[{Kollo and von Rosen(2005)}]{matrix_op}
Kollo, T. and von Rosen, D. (2005), \textit{Advanced Multivariate Statistics
  with Matrices}, Springer.

\bibitem[{Lafit et~al.(2022)Lafit, Nogales, Ruiz, and Zamar}]{lafit2022robust}
Lafit, G.; Nogales, F.; Ruiz, M. and Zamar, R. (2022), \enquote{Robust
  graphical lasso based on multivariate Winsorization,} \textit{Working paper},
  arXiv:2201.03659, 1--33.

\bibitem[{Loh and Tan(2018)}]{loh2018high}
Loh, P.-L. and Tan, X.~L. (2018), \enquote{High-dimensional robust precision
  matrix estimation: Cellwise corruption under $\epsilon $-contamination,}
  \textit{Electronic Journal of Statistics}, 12, 1429--1467.

\bibitem[{{\"O}llerer and Croux(2015)}]{ollerer2015robust}
{\"O}llerer, V. and Croux, C. (2015), \enquote{Robust high-dimensional
  precision matrix estimation,} in \textit{Modern nonparametric, robust and
  multivariate methods}, Springer, pp. 325--350.

\bibitem[{{\"O}llerer et~al.(2015){\"O}llerer, Croux, and
  Alfons}]{ollerer2015influence}
{\"O}llerer, V.; Croux, C. and Alfons, A. (2015), \enquote{The influence
  function of penalized regression estimators,} \textit{Statistics}, 49,
  741--765.

\bibitem[{Raymaekers and Rousseeuw(2021)}]{raymaekers2021fast}
Raymaekers, J. and Rousseeuw, P.~J. (2021), \enquote{Fast robust correlation
  for high-dimensional data,} \textit{Technometrics}, 63, 184--198.

\bibitem[{Ren et~al.(2015)Ren, Sun, Zhang, and Zhou}]{Renasymptotic}
Ren, Z.; Sun, T.; Zhang, C. and Zhou, H. (2015), \enquote{Asymptotic normality
  and optimalities in estimation of large gaussian graphical models,}
  \textit{The Annals of Statistics}, 43, 991--1026.

\bibitem[{Rothman et~al.(2008)Rothman, Bickel, Levina, and
  Zhu}]{rothman2008sparse}
Rothman, A.~J.; Bickel, P.~J.; Levina, E. and Zhu, J. (2008), \enquote{Sparse
  permutation invariant covariance estimation,} \textit{Electronic Journal of
  Statistics}, 2, 494--515.

\bibitem[{Rousseeuw(1984)}]{rousseeuw1984least}
Rousseeuw, P.~J. (1984), \enquote{Least median of squares regression,}
  \textit{Journal of the American statistical association}, 79, 871--880.

\bibitem[{Rousseeuw(1985)}]{rousseeuw1985multivariate}
--- (1985), \enquote{Multivariate estimation with high breakdown point,}
  \textit{Mathematical statistics and applications}, 8, 37.

\bibitem[{Rousseeuw and Croux(1993)}]{rousseeuw1993alternatives}
Rousseeuw, P.~J. and Croux, C. (1993), \enquote{Alternatives to the median
  absolute deviation,} \textit{Journal of the American Statistical
  association}, 88, 1273--1283.

\bibitem[{Rousseeuw and Van~Driessen(1999)}]{rousseeuw1999fast}
Rousseeuw, P.~J. and Van~Driessen, K. (1999), \enquote{A fast algorithm for the
  minimum covariance determinant estimator,} \textit{Technometrics}, 41,
  212--223.

\bibitem[{Spearman(1904)}]{spearman1904}
Spearman, C. (1904), \enquote{General intelligence, objectively determined and
  measured,} \textit{The American Journal of Psychology}, 15, 201--292.

\bibitem[{Srinivasan and Panda(2022)}]{sym_der}
Srinivasan, S. and Panda, N. (2022), \enquote{What is the gradient of a scalar
  function of a symmetric matrix?} \textit{Indian Journal of Pure and Applied
  Mathematics}.

\bibitem[{Tarr et~al.(2016)Tarr, M{\"u}ller, and Weber}]{tarr2016robust}
Tarr, G.; M{\"u}ller, S. and Weber, N.~C. (2016), \enquote{Robust estimation of
  precision matrices under cellwise contamination,} \textit{Computational
  Statistics \& Data Analysis}, 93, 404--420.

\bibitem[{Tukey(1977)}]{tukey1977EDA}
Tukey, J. (1977), \textit{Exploratory Data Analysis}, Massachusetts: Reading
  (Addison-Wesley).

\bibitem[{Yuan and Lin(2007)}]{yuan2007model}
Yuan, M. and Lin, Y. (2007), \enquote{Model selection and estimation in the
  {G}aussian graphical model,} \textit{Biometrika}, 94, 19--35.

\end{thebibliography}

\clearpage
\appendix

\section{Proofs} \label{app:proofs}
We collect all proofs from the paper in this first appendix.
\medskip

\textbf{Proof of Theorem~\ref{thm:IF}:} 
Let $p_m$ be a sequence of convex norms in $C^{\infty}(\mathbb{R}^{p^2})$ which converges to $\|\cdot\|_1^{-\text{diag}}$ in $W^{2,2}(\mathbb{R}^{p^2})$ as $m\rightarrow\infty$ and let $\bT_{ \omega,p_m}(F)$ be defined as in (\ref{Glassosmooth}). 
Without loss of generality, assume that the Hessian matrix $\nabla^2p_m(\cdot)$ is everywhere diagonal\footnote{This is the case for a relaxation proposed in \cite{avellathesis}, page 57-58. The proof of Lemma 1 below shows that the limit is independent of the chosen sequence.}.
Throughout, we write $\boldsymbol{\Omega}=\matvec(\boldsymbol{\omega})$ and $\bT_{{\omega},p_m}(F)=\bT_{{\omega},p_m}$ when there is no ambiguity to the underlying distribution. 
Since the function to be minimised in $\bT_{ \omega,p_m}$ is differentiable, the first order condition writes
$$-\vecmat(\matvec(\bT_{ \omega,p_m}(F))^{-\top})+\vecmat({\bf S}(F))+\lambda\nabla p_m(\bT_{ \omega,p_m}(F))=0.$$
Note that the first term comes from the fact that (see \citealp{matrix_op}, page 134)
$$\frac{d}{d\omega_{i,j}}(-\text{logdet}(\matvec(\boldsymbol{\omega})))=-\frac{1}{\text{det}(\matvec(\boldsymbol{\omega}))}\frac{d}{d\omega_{i,j}}\text{det}(\matvec(\boldsymbol{\omega}))=-\big(\matvec(\boldsymbol{\omega})^{-1}\big)_{j,i},$$
so that $\nabla\big(-\text{logdet}(\matvec(\boldsymbol{\omega}))\big)=-\vecmat(\matvec(\boldsymbol{\omega})^{-\top}).$
Define
$$f(\varepsilon,\boldsymbol{\omega})=-\vecmat(\matvec(\boldsymbol{\omega})^{-1})+\vecmat({\bf S}(F_{\varepsilon, \bz}))+\lambda\nabla p_m(\boldsymbol{\omega}).$$
In particular, it holds $f(0,\bT_{ \omega,p_m}(F))=0$. 
The partial derivatives at $(0,\bT_{ \omega,p_m}(F))$ are given by
\begin{eqnarray*}
\frac{df(\varepsilon,\boldsymbol{\omega})}{d\varepsilon}\Bigr|_{\varepsilon=0}&=& \vecmat(\text{IF}({\bf z};{\bf S}(F),F))\textrm{, \qquad and}\\
 \frac{df(\varepsilon,\boldsymbol{\omega})}{d \boldsymbol{\omega}}\Bigr|_{\boldsymbol{\omega} ={\bT}_{\omega,p_m}}&=&\lambda \nabla^2p_m(T_{\boldsymbol \omega,p_m}) + \bU(T_{\boldsymbol \omega,p_m}),
\end{eqnarray*}
where (see \citealp{matrix_op}, pages 127-131)
$$
\bU(\boldsymbol{\omega}):= \frac{d(-\vecmat(\boldsymbol{\Omega}^{-1}))}{d\vecmat(\boldsymbol{\Omega})}=-\frac{d(\boldsymbol{\Omega}^{-1})}{d\boldsymbol{\Omega}}=\boldsymbol{\Omega}^{-1}\otimes\boldsymbol{\Omega}^{-1}=\matvec(\boldsymbol{\omega})^{-1}\otimes \matvec(\boldsymbol{\omega})^{-1}.
$$
The implicit function theorem yields
$$
\text{IF}(\bz;\bT_{{\omega},p_m},F)=-\big(\lambda\nabla^2p_m(\bT_{ \omega,p_m})+\bU(\bT_{ \omega,p_m})\big)^{-1}\vecmat(\text{IF}(\bz;{\bf S}(F),F)).
$$
Note that this result applies here since the derivative with respect to $\boldsymbol{\omega}$ is nowhere vanishing, a consequence of the positive semi-definiteness of $\nabla^2 p_m(\boldsymbol{\omega})$ and of the positive definiteness of $\bU(\boldsymbol{\omega})$ (see \citealp{sym_der}). 
Taking now the limit in $m$, $ \lim_{m\rightarrow\infty}\text{IF}(\bz;\bT_{{\omega},p_m},F)$ rewrites
\begin{align*}
 &\lim_{m\rightarrow\infty}-(\lambda \nabla^2p_m(\bT_{{\omega},p_m}) + \bU(\bT_{{\omega},p_m}))^{-1}\vecmat(\IF(\bz;{\bf S}(F),F))\\
    &=\lim_{m\rightarrow\infty}-\bD\bD^{-1}(\lambda \nabla^2p_m(\bT_{{\omega},p_m}) + \bU(\bT_{{\omega},p_m}))^{-1} \bD^{-\top}\bD^\top\vecmat(\IF(\bz;{\bf S}(F),F))\\
    &=\lim_{m\rightarrow\infty}-\bD(\bD^\top\lambda \nabla^2p_m(\bT_{{\omega},p_m})\bD + \bD^\top \bU(\bT_{{\omega},p_m})\bD)^{-1} \bD^\top\vecmat(\IF(\bz;{\bf S}(F),F))\\
    &=\lim_{m\rightarrow\infty}-\bD(\lambda \nabla^2p_m(\bD\bT_{{\omega},p_m}) + \bD^\top \bU(\bT_{{\omega},p_m})\bD)^{-1} \bD^\top\vecmat(\IF(\bz;{\bf S}(F),F))\\
&=-\bD\Big(\lim_{m\rightarrow\infty}(\lambda \nabla^2p_m(\bD\bT_{{\omega},p_m}) + \bD^\top \bU(\bT_{{\omega},p_m})\bD)^{-1} \Big)\bD\vecmat(\IF(\bz;{\bf S}(F),F)).
\end{align*}
We now show that the remaining limit converges to
$$\left(
\begin{array}{cc}
(\bD\matvec(\bT_{{\omega}}(F))^{-1} \otimes \matvec(\bT_{{\omega}}(F))^{-1})\bD)^{-1}_{1:s,1:s}  & 0 \\
0&0
\end{array}
\right),$$
which allows to conclude (using ${\bf D}^2={\bf I}_p$).
To show the convergence, partition 
$$\lambda \nabla^2p_m(\bD\bT_{{\omega},p_m}) + \bD^\top \bU(\bT_{{\omega},p_m})\bD=\binom{A_{11}\ A_{12}}{A_{21}\ A_{22}}$$
into blocks of $s$ (resp. $(p^2-s)$) lines/columns. Provided all inverses exist, the inverse is given by
$$
\left(
\begin{array}{cc}
A^{11} & A^{12}\\
A^{21} & A^{22}
\end{array}
\right)
=
\left(
\begin{array}{cc}
(A_{11}-A_{12}A_{22}^{-1}A_{21})^{-1} & -A^{11}A_{12}A_{22}^{-1}\\
-A_{22}^{-1}A_{21}A^{11} & (A_{22}-A_{21}A_{11}^{-1}A_{12})^{-1}
\end{array}
\right).
$$
Notice that, since the inverse function is continuous and the Hessian matrix $ \nabla^2p_m(\bT_{{\omega},p_m})$ is diagonal, it holds $ (\nabla^2p_m(\bD\bT_{\omega}(F)))_{jj}\rightarrow 0$ for $j=1,...,s$ and $ (\nabla^2p_m(\bD\bT_{\omega}(F)))_{jj}$ diverges to $+\infty$ for $j=s+1,...,p^2$.
Therefore, $A_{22}^{-1}\rightarrow 0$, $(A_{22}-A_{21}A_{11}^{-1}A_{12})^{-1}\rightarrow 0$ and $(A_{11}-A_{12}A_{22}^{-1}A_{21})^{-1}\rightarrow (\bD^\top \bU(\bT_{{\omega},p_m})\bD)^{-1}_{1:s,1:s},$ allowing to conclude to the claimed limit.

Note that we formally showed that $ \lim_{m\rightarrow\infty}\text{IF}(\bz;\bT_{{\omega},p_m},F)$ converges to a limit that might be distinct from the influence function should the derivative not exist. 
In our situation, however, the components of the limiting result are either $0$ (showing that the component of the estimated precision matrix stays at zero) or the derivative taken at a positive value (which then is differentiable), which therefore concludes the proof.
\qed

\medskip

\textbf{Proof of Lemma~\ref{lem:pm}}:  
We prove first that $\lim_{m\rightarrow \infty}\bT_{{\omega},p_m}(F)$ is unique. 
Using Lemma 2 in \cite{avellathesis} with $\Lambda_{\lambda}(\boldsymbol{\omega};F,p)= -\text{logdet}(\boldsymbol{\omega}) +\vecmat({\bf S})^\top \boldsymbol{\omega}+\lambda p_m(\boldsymbol{\omega})$ readily gives
\begin{equation}\label{conv}
    \lim_{m\rightarrow\infty}\bT_{{\omega},p_m}(F)=\bT_{{\omega}}(F).
\end{equation}
We can now prove the independence of $ \lim_{m\rightarrow\infty}\IF(\bz;\bT_{{\omega},p_m},F)$ of $p_m$. Let $\{p_m\}_{m\geq1}$ and $\{p'_m\}_{m\geq1}$ be two sequences in $C^{\infty}(\mathbb{R}^{p^2})$ converging to $p$ in $W^{2,2}(\mathbb{R}^{p^2})$. We have
\begin{align*}
     \IF(\bz;\bT_{{\omega},p_m},F)- &\IF(\bz;\bT_{{\omega},p'_m},F)=\big(-(\lambda \nabla^2p_m(\bT_{{\omega},p_m}(F)) 
     + \bU(\bT_{{\omega},p_m}(F))^{-1} \\
     &+(\lambda \nabla^2p'_m(\bT_{{\omega},p'_m}(F)) + \bU(\bT_{{\omega},p'_m}(F))^{-1}\big)\vecmat(\IF(\bz;S(F),F)).
\end{align*}
Since the inverse of a matrix is continuous and $\nabla^2p_m$ and $\nabla^2p'_m$ are sequences in $C^{\infty}(\mathbb{R}^{p^2})$, we find, using ($\ref{conv}$), 
\begin{equation*}
    \lim_{m\rightarrow\infty}(\IF(\bz;\bT_{{\omega},p_m},F)- \IF(\bz;\bT_{{\omega},p'_m},F))=0.
\end{equation*}
Trivially, the convergence becomes uniform if $\sup_{\bz\in\mathbb{R}^p}\|\vecmat(\IF(\bz;S(F),F))\|<\infty$, since it is the only term depending on $\bz$.
\qed 

\medskip

\textbf{Proof of Lemma~\ref{lem:maxdir}:}
Without loss of generality due to the equivariance of ${\bf S}$, we assume throughout that $\mathbb{E}_F[{\bf X}]=0$. It is easy to see that $\IF(\bz;{\bf S},F)=\bz\bz^\top-\boldsymbol{\Sigma}$. 
Hence, since $\textrm{tr}({\bf A}{\bf B}{\bf C})=\textrm{tr}({\bf B}{\bf C}{\bf A})$ and using the spectral decomposition $\boldsymbol{\Omega}={\bf O}\boldsymbol{\Lambda}{\bf O}^\top$, it follows
\begin{eqnarray*}
\|\IF(\bz;\bT_{{\omega}},F)\|^2_F&=&\|\boldsymbol{\Omega} \IF(\bz;{\bf S},F)\boldsymbol{\Omega}\|^2_F=\|\boldsymbol{\Omega} (\bz\bz^\top-\boldsymbol{\Sigma})\boldsymbol{\Omega}\|^2_F=\|\boldsymbol{\Omega}^2 (\bz\bz^\top-\boldsymbol{\Sigma})\|^2_F\\
&=&\|\boldsymbol{\Omega}^2\bz\bz^\top-\boldsymbol{\Omega}\|^2_F=\|\boldsymbol{\Lambda}^2 \boldsymbol{x}\boldsymbol{x}^\top-\boldsymbol{\Lambda}\|^2_F\\
&=&\sum_{i=1}^p \lambda_i^2  +\sum_{i=1}^p x_i^2 (\lambda_i^4-2\lambda_i^3),
\end{eqnarray*}
with $\boldsymbol{x}=(x_1,\dots,x_p)^\top={\bf O}^\top\bz$. 
Since $x_i={\bf v}_i^\top\bz$ and using $\bz=\sum_{j=1}^p\alpha_j {\bf v}_j$, we conclude
$$
\|\IF(\bz;\bT_{{\omega}},F)\|^2_F=\sum_{i=1}^p \lambda_i^2  +\sum_{i=1}^p \alpha_i^2 (\lambda_i^4-2\lambda_i^3)\leq \sum_{i=1}^p \lambda_i^2  + \max_{i\in\{1,...,p\}}(\lambda_i^4-2\lambda_i^3),
$$
which is reached for the eigenvector corresponding to the eigenvalue which maximises $(\lambda^4-2\lambda^3)$. 
Since $g$ is strictly decreasing on the interval $[0,1.5]$ then strictly increasing, the influence function is maximised at either $\lambda_1$ or $\lambda_p$, which completes the proof.
\qed

\section{Robust Correlation Estimators}\label{app:robcor}
To make the paper self-contained, we hereby include an overview of the finite-sample definitions of the different correlation estimators we consider.

First, Spearman’s rank correlation $r_S$ is the sample correlation of the ranks of the observations. Second, Kendall’s correlation between variable $j$ and $k$ is given by 
\begin{equation*}
    r_K =\frac{2}{n(n-1)}\sum_{i<\ell}\textrm{sign}((x_{ij}-x_{\ell j})(x_{ik}-x_{\ell k})).
\end{equation*}
Third, the Gaussian rank correlation between variable $j$ and $k$  is  the sample correlation estimated from the Van Der Waerden scores of the data, as given by
\begin{equation*}
    r_{G}=\frac{\sum_{i=1}^n \Phi^{-1}\left(\dfrac{\text{rank}(x_{ij})}{n+1}\right)\Phi^{-1}\left(\dfrac{\text{rank}(x_{ik})}{n+1}\right)}{\sum_{i=1}^n\left(\Phi^{-1}\left(\dfrac{i}{n+1}\right)\right)^2},
\end{equation*}
where $\text{rank}(x_{ij})$ denotes the rank of $x_{ij}$ among all components of the $j$th variable.
Finally,  the Quadrant correlation between variable $j$ and $k$  is given by
\begin{equation*}
    r_Q=\frac{1}{n}\sum_{i=1}^n\textrm{sign}\left\{(x_{ij}-\textrm{med}_\ell(x_{\ell j}))(x_{ik}-\textrm{med}_\ell(x_{\ell k}))\right\},
\end{equation*}
defined as the difference between, on the one hand, the frequency of the centered observations in the first and third quadrant and, on the other hand, the frequency of the centered observations in the second and fourth quadrant.

~\\
\section{Additional Figures} \label{app:figs_SC}

\begin{figure}[h]
\centering
\includegraphics[width=0.7\linewidth]{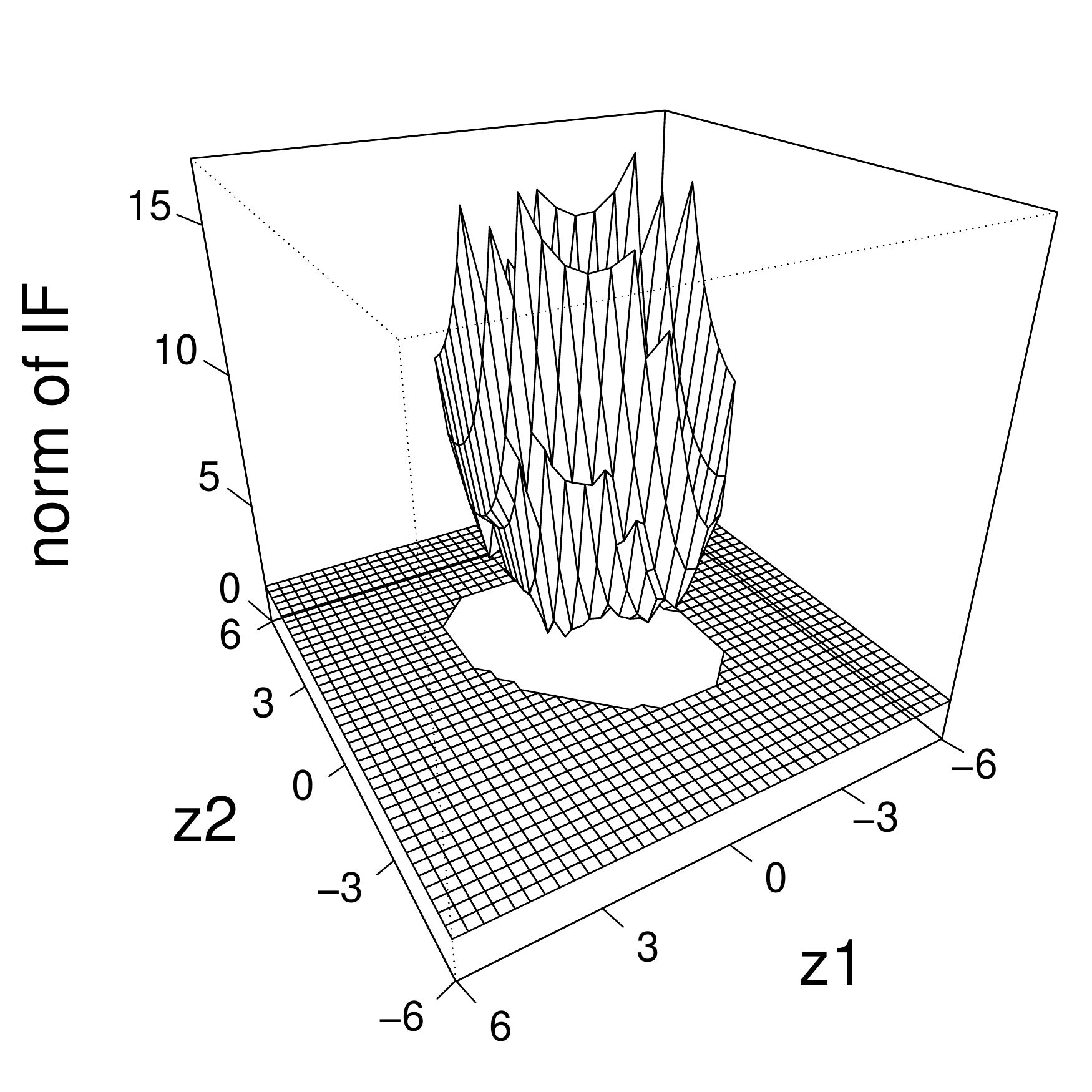}
\caption{Norm of the Glasso influence function based on the reweighted MCD for contamination at $(z_1, z_2, 0)$.}
\label{IF_RMCD}
\end{figure}

\begin{figure}[h!]
  \begin{minipage}[t]{0.32\linewidth}
    \centering
    \includegraphics[width=\linewidth]{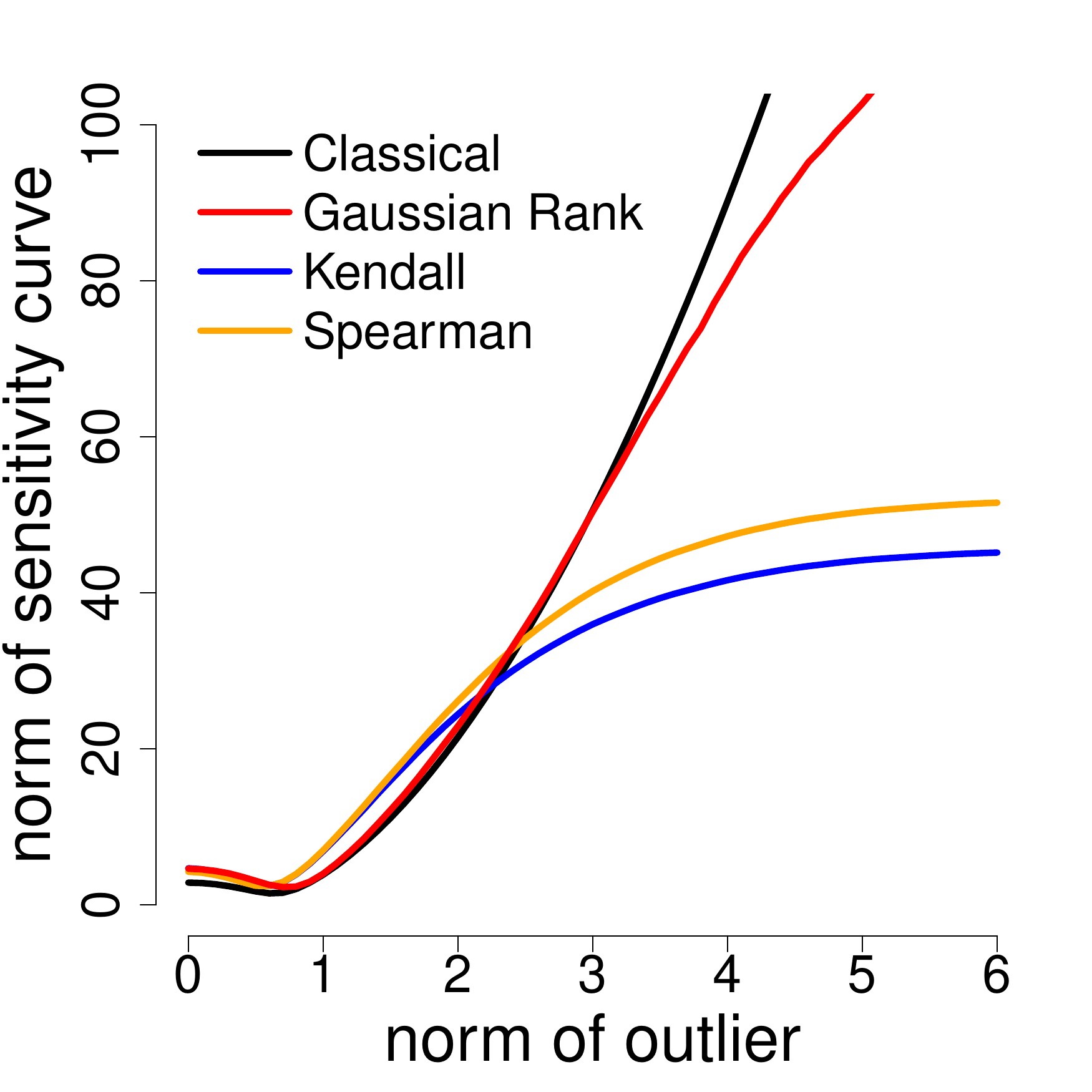} 
  \end{minipage}
  \begin{minipage}[t]{0.32\linewidth}
    \centering
    \includegraphics[width=\linewidth]{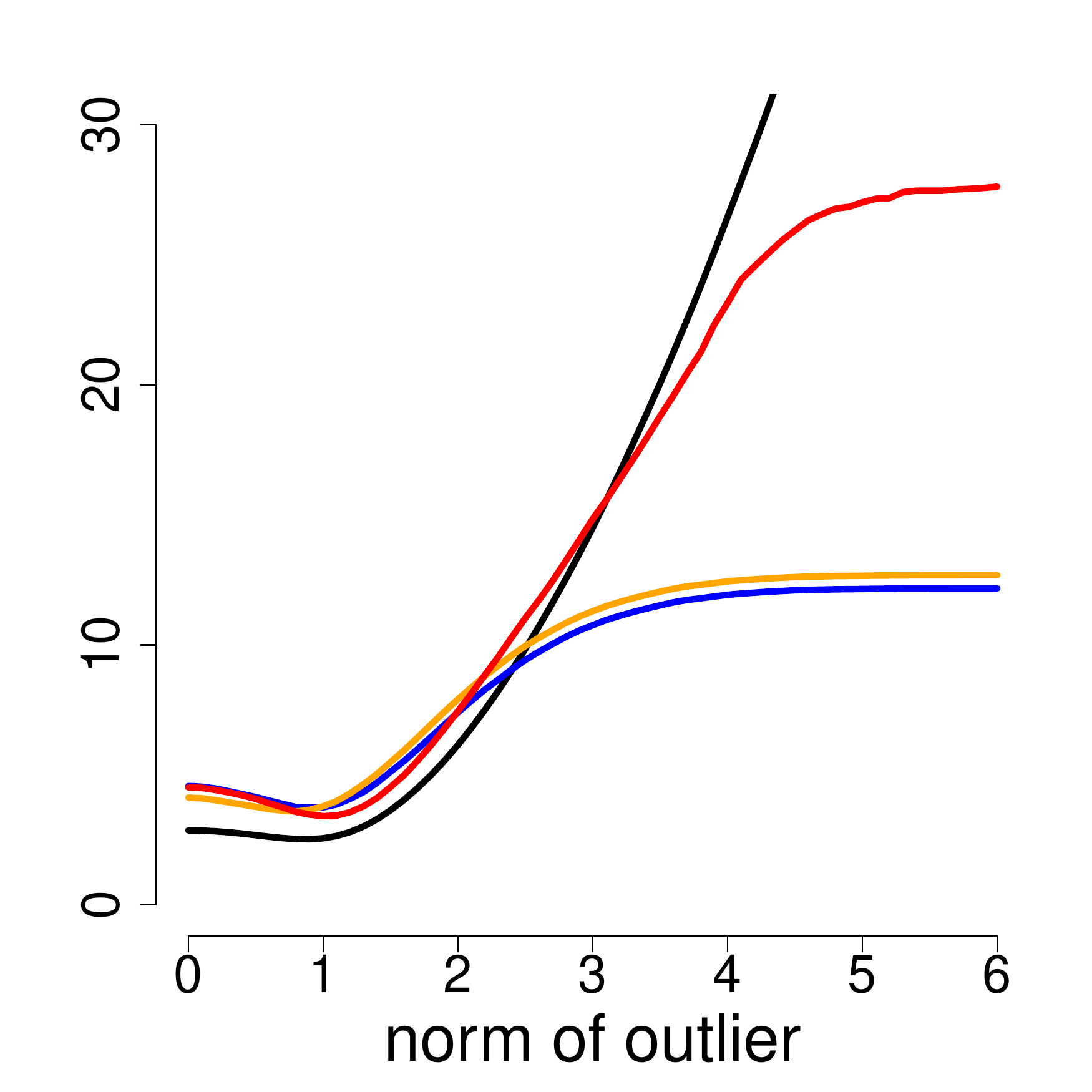} 
   \end{minipage} 
  \begin{minipage}[t]{0.32\linewidth}
    \centering
    \includegraphics[width=\linewidth]{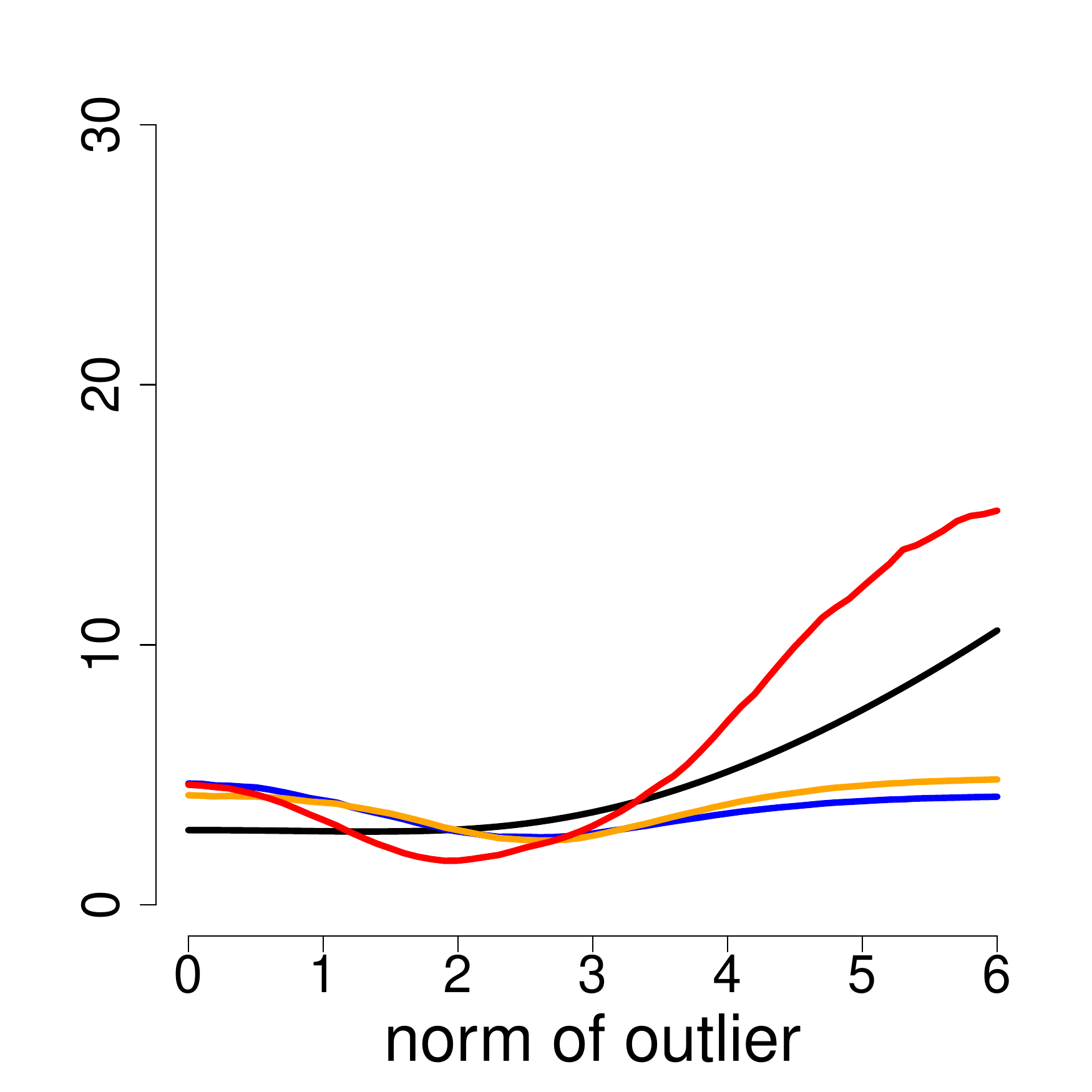} 
   \end{minipage}   
\caption{Norm of the Glasso sensitivity curve based on several covariance matrix estimators for contamination in the direction of the eigenvector corresponding to the first (left), second (middle) and third (right) eigenvalue of $T_\Omega$ based on the true covariance matrix.}
\label{2D_norm_SC} 
\end{figure}

\newpage
\begin{figure}[h]
  \begin{minipage}[b]{0.49\linewidth}
    \centering
    \includegraphics[width=\linewidth]{S_3D_SC_larger_cex.pdf} 
    \caption*{Classical}
  \end{minipage}
  \begin{minipage}[b]{0.49\linewidth}
    \centering
    \includegraphics[width=\linewidth]{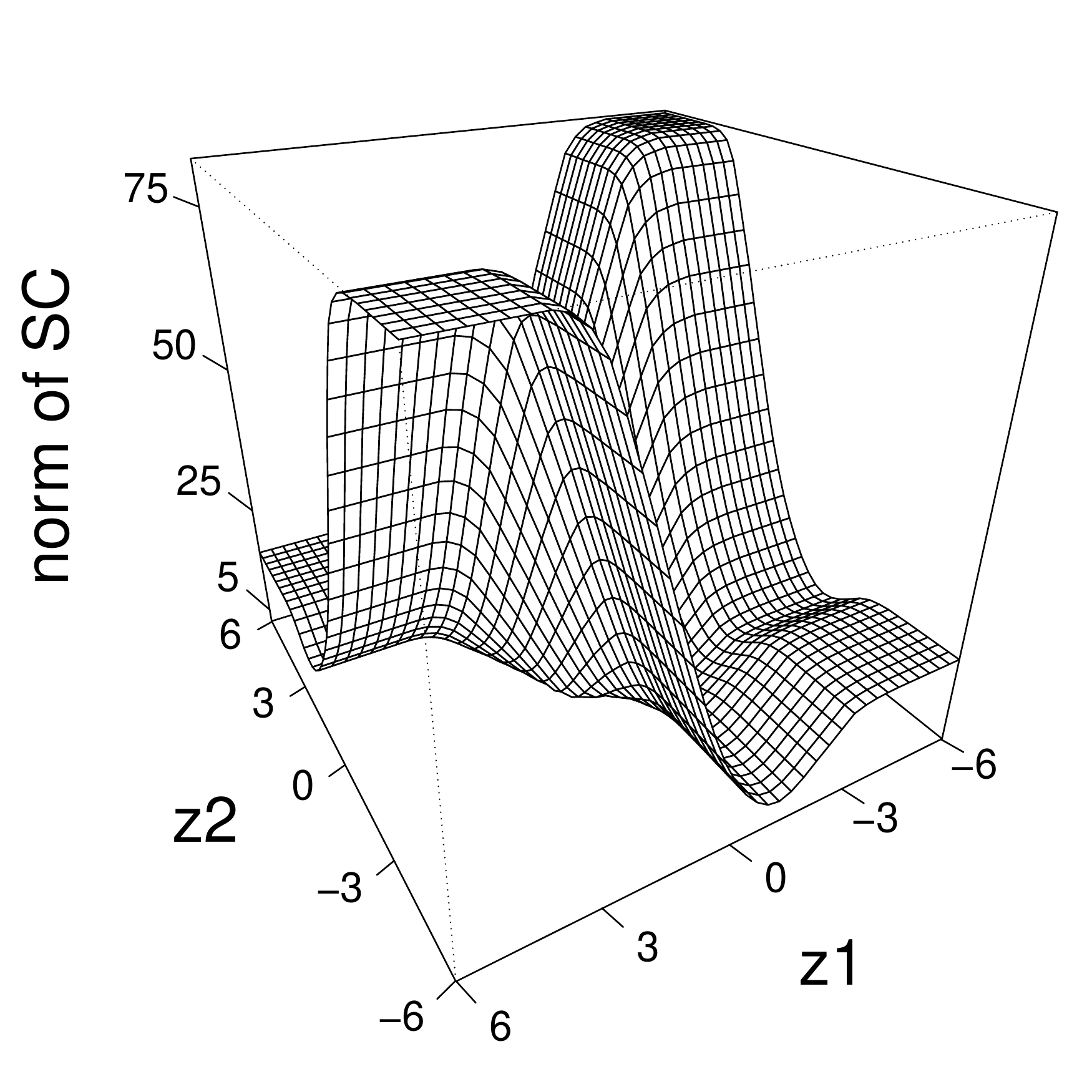} 
    \caption*{Gaussian rank}
   \end{minipage}   
  
   \begin{minipage}[b]{0.49\linewidth}
    \centering
    \includegraphics[width=\linewidth]{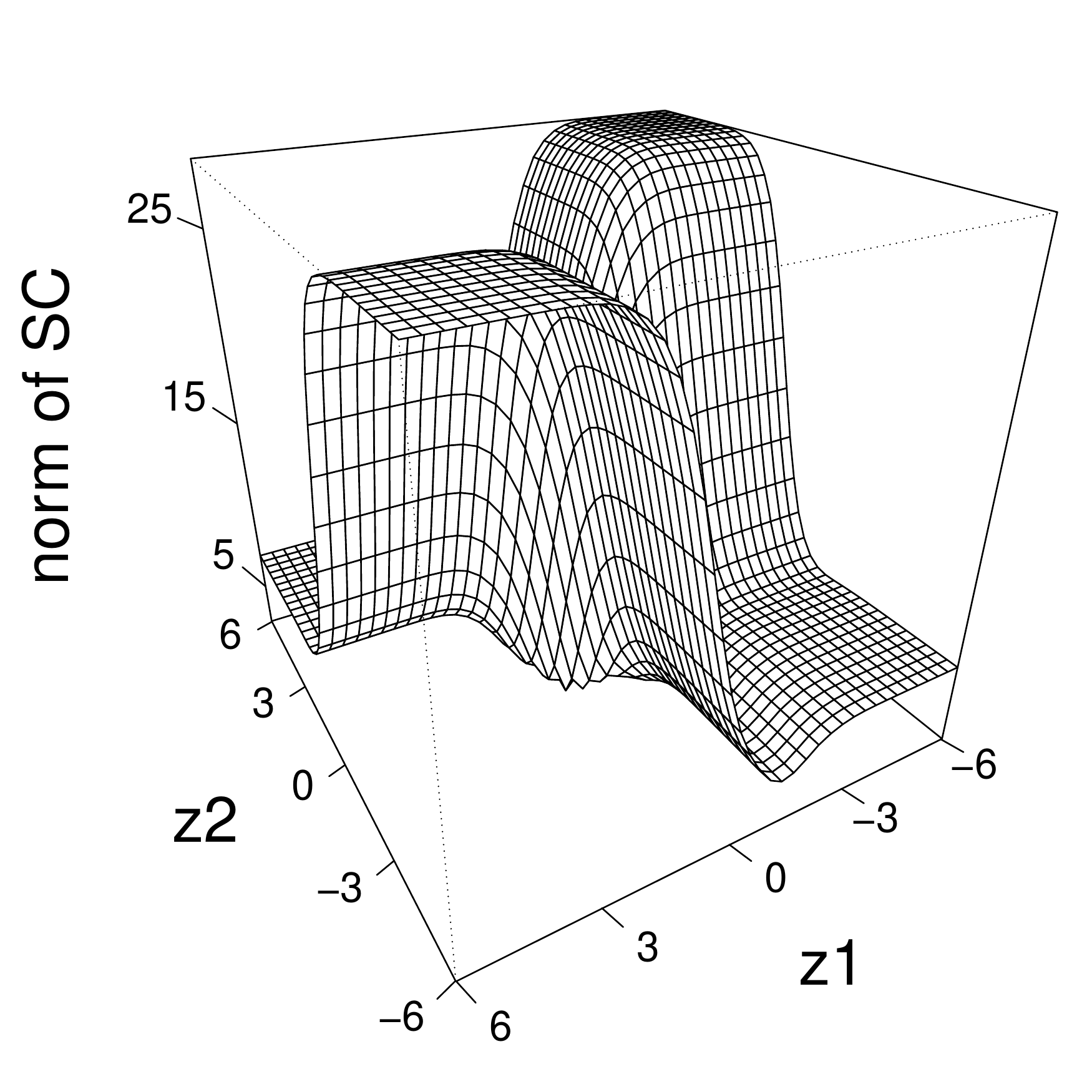} 
    \caption*{Kendall}
  \end{minipage}
     \begin{minipage}[b]{0.49\linewidth}
    \centering
    \includegraphics[width=\linewidth]{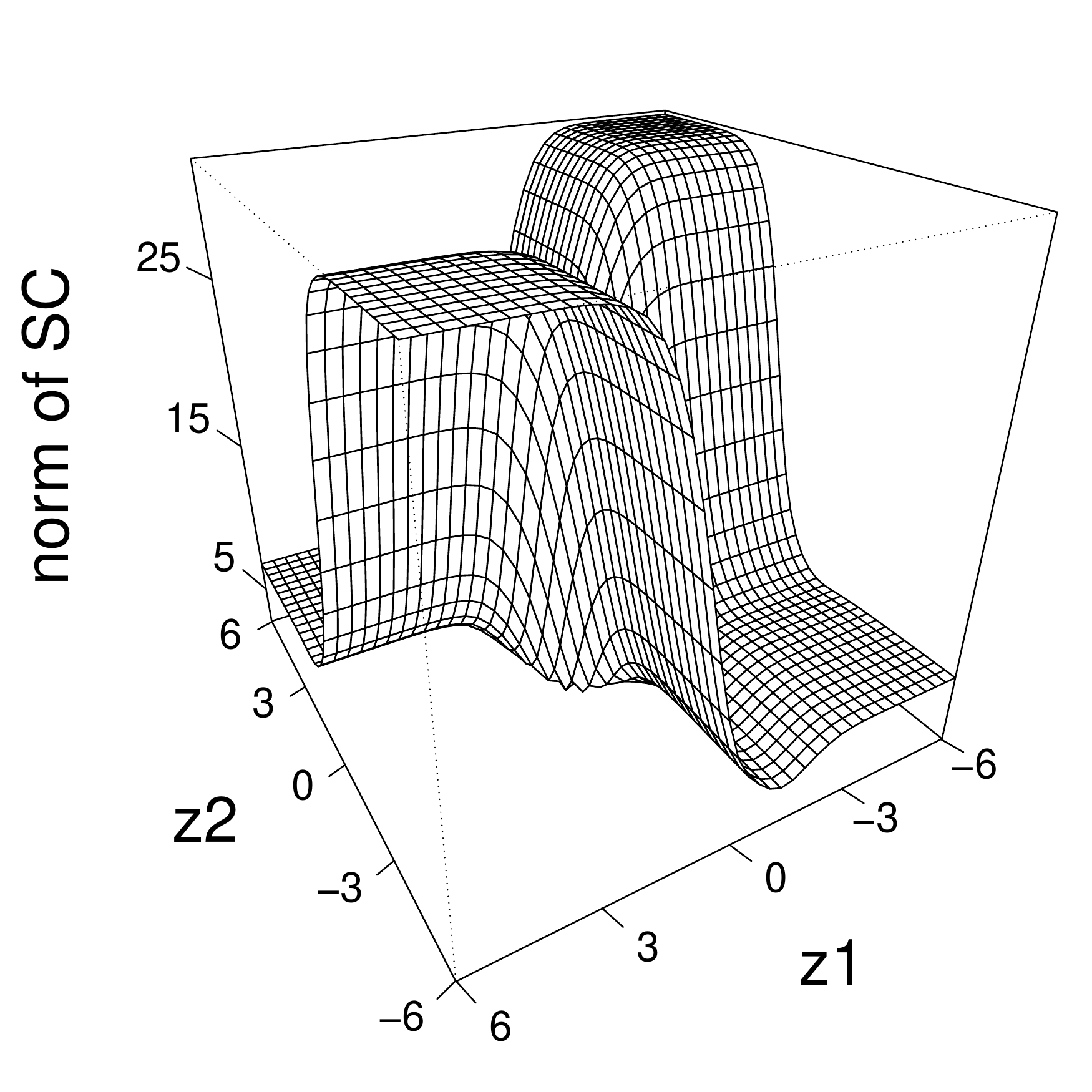} 
    \caption*{Spearman}
  \end{minipage}
  \caption{Norm of the Glasso sensitivity curve based on several covariance matrix estimators for contamination at $(z_1, z_2, 0)$.}
  \label{3DSC}
\end{figure}

\end{document}